\documentclass[epj]{svjour}
\usepackage{graphicx}
\usepackage{braket}
\usepackage{amsfonts}
\usepackage{amssymb}
\usepackage{amsmath}
\usepackage{multirow}
\usepackage{hyperref}
\usepackage{breakurl}
\usepackage[numbers, sort&compress]{natbib}
\newcommand*{\doi}[1]{\href{http://dx.doi.org/#1}{doi: #1}}
\begin{document}
\title{The density matrix renormalization group for ab initio quantum chemistry}
\author{Sebastian {Wouters} \and Dimitri {Van Neck}}
%
%
\institute{Center for Molecular Modelling, Ghent University, Technologiepark 903, 9052 Zwijnaarde, Belgium}
\date{Received: date / Revised version: date}
%
\abstract{
During the past 15 years, the density matrix renormalization group (DMRG) has become increasingly important for \textit{ab initio} quantum chemistry. Its underlying wavefunction ansatz, the matrix product state (MPS), is a low-rank decomposition of the full configuration interaction tensor. The virtual dimension of the MPS, the rank of the decomposition, controls the size of the corner of the many-body Hilbert space that can be reached with the ansatz. This parameter can be systematically increased until numerical convergence is reached. The MPS ansatz naturally captures exponentially decaying correlation functions. Therefore DMRG works extremely well for noncritical one-dimensional systems. The active orbital spaces in quantum chemistry are however often far from one-dimensional, and relatively large virtual dimensions are required to use DMRG for \textit{ab initio} quantum chemistry (QC-DMRG). The QC-DMRG algorithm, its computational cost, and its properties are discussed. Two important aspects to reduce the computational cost are given special attention: the orbital choice and ordering, and the exploitation of the symmetry group of the Hamiltonian. With these considerations, the QC-DMRG algorithm allows to find numerically exact solutions in active spaces of up to 40 electrons in 40 orbitals.
} 
\maketitle

\section{Introduction}
At the basis of \textit{ab initio} quantum chemistry lies Hartree-Fock (HF) theory \cite{PSP.1733252, PhysRev.32.339, Fock1}. In HF theory, a single Slater determinant (SD) is optimized by finding the set orbitals which minimize its energy expectation value. The occupancy of the HF orbitals is definite: occupied orbitals are filled with probability 1, and virtual orbitals are empty with probability 1. The exact ground state is a linear combination over all possible Slater determinants. The difference in energy between the HF solution and the exact ground state is the correlation energy. This energy is often (somewhat ambiguously) divided into two contributions: static and dynamic correlation \cite{helgaker2}. When near-degeneracies between determinants occur, and more than one determinant is needed to describe the qualitative behaviour of a molecule, it is said to have static correlation. This type of correlation often arises in transition metal complexes or $\pi$-conjugated systems, as well as for geometries far from equilibrium. It is typically resolved with only a few determinants. The Coulomb repulsion results in a small nonzero occupancy of many virtual HF orbitals in the true ground state. This effect is called dynamic correlation, and it constitutes the remainder of the energy gap.

All static and dynamic correlation can in principle be retrieved at HF cost with density functional theory (DFT). Hohenberg and Kohn have shown that the electron density provides sufficient information to determine all ground state properties, and that there exists a unique universal functional of the electron density which can be used to obtain the exact ground state density \cite{PhysRev.136.B864}. Kohn and Sham rewrote the universal functional as the sum of the kinetic energy of a noninteracting system and an exchange-correlation functional \cite{PhysRev.140.A1133}. This allows to represent the electron density by means of  the Kohn-Sham Slater determinant, which immediately ensures correct N-representability. Unfortunately, the universal functional is unknown. Many approximate semi-empirical exchange-correlation functionals of various complexity have been proposed. Because the exact exchange-correlation functional is unknown, not all correlation is retrieved with DFT. For single-reference systems, for which the exact solution is dominated by a single SD, DFT is good in capturing dynamic correlation. For multireference (MR) systems, DFT fails to retrieve static correlation \cite{BeckeQuote}.

Dynamic correlation can also be captured with \textit{ab initio} post-HF methods. These start from the optimized HF orbitals and the corresponding SD, and build in dynamic correlation on top of the single SD reference. Commonly known are M\o{}ller-Plesset (Rayleigh-Schr\"odinger) perturbation theory \cite{PhysRev.46.618}, the configuration interaction (CI) expansion \cite{PhysRev.34.1293, PhysRev.36.1121}, and coupled cluster (CC) theory \cite{Coester1958421, Coester1960477, CizekCC}. These methods are truncated in their perturbation or expansion order. An important property of wavefunctions is size-consistency: the fact that for two noninteracting subsystems, the compound wavefunction should be multiplicatively separable and the total energy additively separable. CI with $N$ excitations is not size-consistent if there are more than $N$ electrons in the compound system, whereas CC is always size-consistent because of its exponential wavefunction ansatz \cite{helgaker2}. Because these post-HF methods start from a single SD reference, they have difficulty building in static correlation. Mostly, very large expansion orders are required to retrieve static correlation.

It is therefore better to resort to MR methods for systems with pronounced static correlation. For such systems, the subset of important orbitals (the active space), in which the occupation changes over the dominant determinants, is often rather small. This allows for a particular MR solution method: the complete active space (CAS) self-consistent field (SCF) method \cite{Roos1, Roos2, Roos3}. From the HF solution, a subset of occupied and virtual orbitals is selected as active space. While the remaining occupied and virtual orbitals are kept frozen at HF level, the electronic structure in the active space is solved exactly (the CAS-part). Subsequently, the occupied, active, and virtual spaces are rotated to further minimize the energy. This two-step cycle, which is sometimes implemented together, is repeated until convergence is reached (the SCF-part). CASSCF resolves the static correlation in the system. Dynamic correlation can be built in on top of the CASSCF reference wavefunction by perturbation theory (CASPT2) \cite{CASPT2-1,CASPT2-2}, a CI expansion (MRCI or CASCI) \cite{MRCIfirst, MRCIsecond, CASCI-1, CASCI-2, CASCI-3}, or CC theory (MRCC or CASCC) \cite{MRCC,Stolarczyk19941}. For the latter, approximate schemes such as canonical transformation (CT) theory \cite{CT-first} are often used.

Because the many-body Hilbert space grows exponentially with the number of single-particle states, only small active spaces, of up to 18 electrons in 18 orbitals, can be treated in the CAS-part. In 1999, the density matrix renormalization group (DMRG) was introduced in \textit{ab initio} quantum chemistry (QC) \cite{WhiteQCDMRG}. This MR method allows to find numerically exact solutions in significantly larger active spaces, of up to 40 electrons in 40 orbitals.

\section{Matrix product states} \label{sec1p4-in-chap1}
The electronic Hamiltonian can be written in second quantization as
\begin{eqnarray}
\hat{H} & = & E_0 + \sum\limits_{ij} t_{ij} \sum\limits_{\sigma} \hat{a}_{i\sigma}^{\dagger} \hat{a}_{j\sigma} \nonumber \\
& + & \frac{1}{2} \sum\limits_{ijkl} v_{ij;kl} \sum\limits_{\sigma\tau} \hat{a}_{i\sigma}^{\dagger} \hat{a}_{j\tau} ^{\dagger} \hat{a}_{l\tau} \hat{a}_{k\sigma}. \label{QC-ham}
\end{eqnarray}
The Latin letters denote spatial orbitals and the Greek letters electron spin projections. The $t_{ij}$ and $v_{ij;kl}$ are the one- and two-electron integrals, respectively. In the occupation number representation, the basis states of the many-body Hilbert space are
\begin{eqnarray}
& \ket{n_{1\uparrow} n_{1\downarrow} ... n_{L \uparrow} n_{L \downarrow}} = \nonumber \\
& \left( \hat{a}^{\dagger}_{1 \uparrow} \right)^{n_{1\uparrow}} \left( \hat{a}^{\dagger}_{1 \downarrow} \right)^{n_{1\downarrow}}  ...  \left( \hat{a}^{\dagger}_{L\uparrow} \right)^{n_{L \uparrow}} \left( \hat{a}^{\dagger}_{L\downarrow} \right)^{n_{L \downarrow}}  \ket{-}. \label{occupation-number-representation-spin-orbs}
\end{eqnarray}

The symmetry group of the Hamiltonian \eqref{QC-ham} is $\mathsf{SU(2)} \otimes \mathsf{U(1)} \otimes \mathsf{P}$, or total electronic spin, particle-number, and molecular point group symmetry. By defining the operators
{\allowdisplaybreaks
\begin{eqnarray}
\hat{S}^{+} & = & \sum\limits_i \hat{a}_{i \uparrow}^{\dagger} \hat{a}_{i \downarrow}, \\
\hat{S}^{-} & = & \left( \hat{S}^{+} \right)^{\dagger} = \sum\limits_i \hat{a}_{i \downarrow}^{\dagger} \hat{a}_{i \uparrow}, \\
\hat{S}^{z} & = & \frac{1}{2} \sum\limits_i \left( \hat{a}_{i \uparrow}^{\dagger} \hat{a}_{i \uparrow} - \hat{a}_{i \downarrow}^{\dagger} \hat{a}_{i \downarrow} \right), \\
\hat{N} & = & \sum\limits_i \left( \hat{a}_{i \uparrow}^{\dagger} \hat{a}_{i \uparrow} + \hat{a}_{i \downarrow}^{\dagger} \hat{a}_{i \downarrow} \right), \\
\hat{S}^2 & = & \frac{\hat{S}^+ \hat{S}^- + \hat{S}^- \hat{S}^+}{2} + \hat{S}^z \hat{S}^z,
\end{eqnarray}
}it can be easily checked that $\hat{H}$, $\hat{S}^2$, $\hat{S}^z$, and $\hat{N}$ form a set of commuting observables. This constitutes the $\mathsf{SU(2)}$ total electronic spin and $\mathsf{U(1)}$ particle-number symmetries. For fixed particle number $N$, Eq. \eqref{QC-ham} can also be written as
\begin{equation}
\hat{H} = E_0 + \frac{1}{2} \sum\limits_{ijkl} h_{ij;kl} \sum\limits_{\sigma\tau} \hat{a}_{i\sigma}^{\dagger} \hat{a}_{j\tau} ^{\dagger} \hat{a}_{l\tau} \hat{a}_{k\sigma}, \label{QC-ham-2} \\
\end{equation}
with
\begin{equation}
h_{ij;kl} = v_{ij;kl} + \frac{1}{N-1} \left( t_{ik} \delta_{j,l} + t_{jl} \delta_{i,k} \right).
\end{equation}

The molecular point group symmetry $\mathsf{P}$ consists of the rotations, reflections, and inversions which leave the external potential due to the nuclei invariant. These symmetry operations map nuclei with equal charges onto each other. The point group symmetry has implications for the spatial orbitals. Linear combinations of the single-particle basis functions can be constructed which transform according to a particular row of a particular irreducible representation (irrep) of $\mathsf{P}$ \cite{BookCornwell}. As the Hamiltonian transforms according to the trivial irrep $I_0$ of $\mathsf{P}$, $h_{ij;kl}$ can only be nonzero if the reductions of $I_i \otimes I_j$ and $I_k \otimes I_l$ have at least one irrep in common. Most molecular electronic structure programs make use of the abelian point groups with real-valued character tables.

\begin{figure}[t!]
\centering
\includegraphics[width=0.22\textwidth]{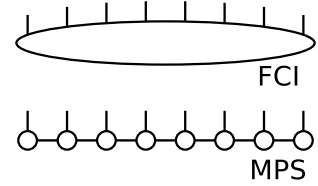}
\caption{\label{FCItoMPS-plot} Tensors are represented by circles, physical indices by open lines, and virtual indices by connected lines. The MPS graph hence represents how the contracted matrix product decomposes the FCI tensor.}
\end{figure}

An eigenstate of the Hamiltonian \eqref{QC-ham-2} can be written as
\begin{eqnarray}
\ket{\Psi} & = & \sum_{\{n_{j\sigma} \}} C^{n_{1\uparrow} n_{1\downarrow} n_{2\uparrow} n_{2\downarrow}  ... n_{L\uparrow} n_{L\downarrow}} \nonumber \\
& & \ket{n_{1\uparrow} n_{1\downarrow} n_{2\uparrow} n_{2\downarrow}  ... n_{L\uparrow} n_{L\downarrow}}. \label{asdfhasdgjhsajcucahsgasgsjkkskkkk}
\end{eqnarray}
The size of the full CI (FCI) tensor grows as $4^L$, exponentially fast with $L$. This tensor can be exactly decomposed by a singular value decomposition (SVD) as follows:
{\allowdisplaybreaks
\begin{eqnarray}
& C^{n_{1\uparrow} n_{1\downarrow} n_{2\uparrow} n_{2\downarrow} ... n_{L\uparrow} n_{L\downarrow}} = \nonumber \\
& C_{(n_{1\uparrow} n_{1\downarrow});(n_{2\uparrow} n_{2\downarrow} ... n_{L\uparrow} n_{L\downarrow})} = \nonumber \\
& \sum\limits_{\alpha_1} U[1]_{(n_{1\uparrow} n_{1\downarrow}); \alpha_1 } s[1]_{\alpha_1} V[1]_{\alpha_1 ; (n_{2\uparrow} n_{2\downarrow} ... n_{L\uparrow} n_{L\downarrow})}.
\end{eqnarray}
}Define
\begin{equation}
A[1]^{n_{1\uparrow}n_{1\downarrow}}_{\alpha_1} = U[1]_{(n_{1\uparrow} n_{1\downarrow}); \alpha_1 } s[1]_{\alpha_1},
\end{equation}
and decompose the right unitary $V[1]$ again with an SVD as follows:
\begin{eqnarray}
& V[1]_{\alpha_1 ; (n_{2\uparrow} n_{2\downarrow} n_{3\uparrow} n_{3\downarrow}  ... n_{L\uparrow} n_{L\downarrow})} = \nonumber \\
& V[1]_{(\alpha_1 n_{2\uparrow} n_{2\downarrow});(n_{3\uparrow} n_{3\downarrow}  ... n_{L\uparrow} n_{L\downarrow})} = \nonumber \\
& \sum\limits_{\alpha_2} U[2]_{(\alpha_1 n_{2\uparrow} n_{2\downarrow}); \alpha_2 } s[2]_{\alpha_2} V[2]_{\alpha_2 ; (n_{3\uparrow} n_{3\downarrow}  ... n_{L\uparrow} n_{L\downarrow})}.
\end{eqnarray}
Define
\begin{equation}
A[2]^{n_{2\uparrow}n_{2\downarrow}}_{\alpha_1;\alpha_2} = U[2]_{(\alpha_1 n_{2\uparrow} n_{2\downarrow}); \alpha_2 } s[2]_{\alpha_2}.
\end{equation}
Continue by successively decomposing the right unitaries $V[k]$. In this way, the FCI tensor can be exactly rewritten as the following contracted matrix product:
\begin{eqnarray}
& C^{n_{1\uparrow} n_{1\downarrow} n_{2\uparrow} n_{2\downarrow} n_{3\uparrow} n_{3\downarrow}  ... n_{L\uparrow} n_{L\downarrow}} = \nonumber \\
& \sum\limits_{\{ \alpha_k \}} A[1]^{n_{1\uparrow}n_{1\downarrow}}_{\alpha_1} A[2]^{n_{2\uparrow}n_{2\downarrow}}_{\alpha_1 ; \alpha_2} A[3]^{n_{3\uparrow}n_{3\downarrow}}_{\alpha_2 ; \alpha_3} ... A[L]^{n_{L\uparrow}n_{L\downarrow}}_{\alpha_{L-1}}, \label{MPS-A-tensor}
\end{eqnarray}
which is graphically represented in Fig. \ref{FCItoMPS-plot}. Except for the first and last orbital (or \textit{site}), Eq. \eqref{MPS-A-tensor} introduces a rank-3 tensor per site. One of its indices corresponds to the physical index $n_{i\uparrow}n_{i\downarrow}$, the other two to the \textit{virtual} or \textit{bond} indices $\alpha_{i-1}$ and $\alpha_i$. In Fig. \ref{FCItoMPS-plot}, tensors are represented by circles, physical indices by open lines, and virtual indices by connected lines. The graph hence represents how the contracted matrix product decomposes the FCI tensor. Since no assumptions are made about the FCI tensor, the dimension of the indices $\{\alpha_k\}$ has to grow exponentially towards the middle of this contracted product:
\begin{equation}
 \text{dim}\left( \alpha_j \right) = \min \left( 4^j , 4^{L-j} \right).
\end{equation}
This is solely due to the increasing matrix dimensions in the successive SVDs. Instead of variationally optimizing over the FCI tensor, one may as well optimize over the tensors of its decomposition \eqref{MPS-A-tensor}. To make Eq. \eqref{MPS-A-tensor} of practical use, its dimensions can be truncated:
\begin{equation}
 \text{dim}\left( \alpha_j \right) = \min \left( 4^j , 4^{L-j}, D \right). \label{MPSapproxSizes}
\end{equation}
The corresponding ansatz is called a matrix product state (MPS) with open boundary conditions. The truncation dimension $D$ is called the \textit{bond} or \textit{virtual} dimension. The MPS ansatz can be optimized by the DMRG algorithm \cite{PhysRevLett.69.2863, PhysRevB.48.10345, WhiteQCDMRG}, yielding a variational upper bound for the ground state energy.

DMRG was invented in 1992 by White in the field of condensed matter theory \cite{PhysRevLett.69.2863}. \"Ostlund and Rommer discovered in 1995 its underlying variational ansatz, the MPS \cite{PhysRevLett.75.3537, PhysRevB.55.2164}. The discovery of the MPS ansatz allowed to understand DMRG by means of quantum information theory. The area law for one-dimensional quantum systems, see section \ref{entanglement-section}, was proven by Hastings in 2007 \cite{1742-5468-2007-08-P08024}, and constitutes a hard proof that an MPS is very efficient in representing the ground state of noncritical one-dimensional quantum systems.

The MPS ansatz was in fact discovered earlier, under various names. Nishino found that they were used in statistical physics as a variational optimization technique \cite{NishinoHistory}: in 1941 by Kramers and Wannier \cite{PhysRev.60.263} and in 1968 by Baxter \cite{Baxter}. Nightingale and Bl\"ote recycled Baxter's ansatz in 1986 to approximate quantum eigenstates \cite{PhysRevB.33.659}. In 1987, Affleck, Kennedy, Lieb and Tasaki constructed the exact valence-bond ground state of a particular next-nearest-neighbour spin chain \cite{PhysRevLett.59.799}. They obtained an MPS with bond dimension 2. In mathematics, the translationally invariant valence-bond state is known as a finitely correlated state \cite{0295-5075-10-7-005, FCSmath}, and in the context of information compression, an MPS is known as a tensor train \cite{TensorTrains, TensorTrainsNMRspinSystem}.

The concept of a renormalization group was first used in quantum electrodynamics. The coarse-grained view of a point-like electron breaks down at small distance scales (or large energy scales). The electron itself consists of electrons, positrons, and photons. The mass and charge contributions from this fine structure lead to infinities. These were successfully resolved by Tomonaga, Schwinger, and Feynman \cite{Tomonaga01081946, PhysRev.73.416, PhysRev.74.1439, PhysRev.76.769, PhysRev.76.749}. Later, Wilson used a numerical renormalization group (NRG) to solve the long-standing Kondo problem \cite{RevModPhys.47.773}. He turned the coupling of the impurity to the conduction band into a half-infinite lattice problem by discretizing the conduction band in momentum space. For increasing lattice sizes, only the lowest energy states are kept at each renormalization step. These are sufficient to study the low-temperature thermodynamics of the impurity system. Although very successful for impurity systems, NRG fails for real-space lattice systems such as the discretized particle-in-a-box, spin-lattice, and Hubbard models. For these systems, the low energy states of a small subsystem are often irrelevant for the ground state of the total system \cite{PhysRevLett.68.3487}. Consider for example the ground state of the particle-in-a-box problem. By concatenating the solution of two smaller sized boxes, an unphysical node is introduced in the approximation of the ground state of the larger problem. It was White who pointed out this problem and resolved it with his DMRG method \cite{PhysRevLett.69.2863}. Instead of selecting the degrees of freedom with lowest energy, the most relevant degrees of freedom should be selected.

\section{Entanglement and the von Neumann entropy} \label{entanglement-section}
This section attempts to clarify the broader context of DMRG. A brief introduction to quantum entanglement, the von Neumann entropy, and the area law is given.

\begin{figure}[t!]
\centering
\includegraphics[width=0.37\textwidth]{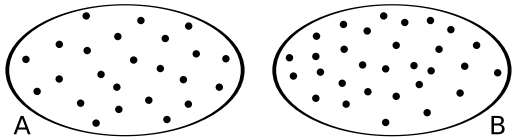}
\caption{\label{Bipartition-plot} Bipartition of the $L$ single-particle states.}
\end{figure}
Consider the bipartition of $L$ orthonormal single-particle states into two subsystems $A$ and $B$ in Fig. \ref{Bipartition-plot}. Suppose $\{ \ket{A_i} \}$ and $\{ \ket{B_j} \}$ are the orthonormal basis states of the many-body Hilbert spaces of resp.\ subsystem $A$ and $B$. The Hilbert space of the composite system is spanned by the product space $\{ \ket{A_i} \} \otimes \{ \ket{B_j} \}$, and a general quantum many-body state $\ket{\Psi}$ of the composite system can be written as
\begin{equation}
\ket{\Psi} = \sum_{ij} C_{ij} \ket{A_i} \ket{B_j}.\label{chap2-eq21-genrealstate}
\end{equation}
The Schmidt decomposition of $\ket{\Psi}$ is obtained by performing an SVD on $C_{ij}$ and by rotating the orthonormal bases $\{ \ket{A_i} \}$ and $\{ \ket{B_j} \}$ with the unitary matrices:
\begin{eqnarray}
\ket{\Psi} & = & \sum_{ij} C_{ij} \ket{A_i} \ket{B_j} = \sum_{ijk} U_{ik} \sigma_k V_{kj} \ket{A_i} \ket{B_j} \nonumber \\
& = & \sum\limits_k \sigma_k \ket{\widetilde{A}_k} \ket{\widetilde{B}_k} .\label{Schmidt-decomp}
\end{eqnarray}
For normalized $\ket{\Psi}$:
\begin{equation}
\braket{\Psi \mid \Psi} = \sum\limits_k \sigma_k^2 = 1.
\end{equation}
For the given bipartition, one is sometimes interested in the optimal approximation $\ket{\widetilde{\Psi}}$ of $\ket{\Psi}$ in a least squares sense $\| \ket{\widetilde{\Psi}} - \ket{\Psi} \|_2$. It can be shown that the optimal approximation, with a smaller number of terms in the summation \eqref{chap2-eq21-genrealstate}, is obtained by keeping the states with the largest Schmidt numbers $\sigma_k$ in Eq. \eqref{Schmidt-decomp}. This fact will be of key importance for the DMRG algorithm (see section \ref{subsec-micro-iteraions}).

In classical theories, the sum over $k$ can contain only one nonzero value $\sigma_k$. A measurement in subsystem $A$ then does not influence the outcome in subsystem $B$, and the two subsystems are not entangled. In quantum theories, the sum over $k$ can contain many nonzero values $\sigma_k$. State $\ket{\widetilde{A}_k}$ in subsytem $A$ occurs with probability $\sigma_{k}^2$, as can be observed from the reduced density matrix (RDM) of subsystem $A$:
\begin{eqnarray}
\hat{\rho}^A & = & \text{Tr}_{B} \ket{\Psi} \bra{\Psi} = \sum\limits_{j} \braket{B_j \mid \Psi} \braket{\Psi \mid B_j} \nonumber \\
& = & \sum\limits_{ijl} \ket{A_i} C_{ij} C^{\dagger}_{jl} \bra{A_l} = \sum\limits_{k} \ket{\widetilde{A}_k} \sigma_k^2 \bra{\widetilde{A}_k}.
\end{eqnarray}
Analogously the RDM of subsystem $B$ can be constructed:
\begin{equation}
\hat{\rho}^B = \sum\limits_{k} \ket{\widetilde{B}_k} \sigma_k^2 \bra{\widetilde{B}_k}.
\end{equation}

From \eqref{Schmidt-decomp}, it follows that the measurement of $\ket{\widetilde{A}_k}$ in subsystem $A$ implies the measurement of $\ket{\widetilde{B}_k}$ in subsystem $B$ with probability 1. Measurements in $A$ and $B$ are hence not independent, and the two subsystems are said to be entangled.

Consider for example two singly occupied orbitals $A$ and $B$ in the spin-0 singlet state:
\begin{equation}
\ket{\Psi} = \frac{\ket{\uparrow_A \downarrow_B} - \ket{\downarrow_A \uparrow_B}}{\sqrt{2}}.
\end{equation}
The measurements of the spin projections of the electrons are not independent. Each possible spin projection of the electron in $A$ can be measured with probability $\frac{1}{2}$, but the simultaneous measurement of both spin projections will always yield
\begin{equation}
\braket{\Psi \mid \hat{S}^z_A \hat{S}^z_B \mid \Psi} = -\frac{1}{4}
\end{equation}
with probability 1.

The RDMs $\hat{\rho}^A$ and $\hat{\rho}^B$ allow to define the von Neumann entanglement entropy \cite{Neumann1927}:
\begin{eqnarray}
S_{A \mid B} & = & - \text{Tr}_A ~ \hat{\rho}^A \ln \hat{\rho}^A = - \text{Tr}_B ~ \hat{\rho}^B \ln \hat{\rho}^B \nonumber \\
& = & - \sum\limits_k  \sigma_k^2 \ln \sigma_k^2. \label{von-neumann-entropy-eq}
\end{eqnarray}
This quantum analogue of the Shannon entropy is a measure of how entangled subsystems $A$ and $B$ are. If they are not entangled, $\sigma_1=1$ and $\sigma_k = 0$ for $k \geq 2$, which implies $S_{A \mid B} = 0$. If they are maximally entangled, $\sigma_k = \sigma_l$ for all $k$ and $l$, which implies $S_{A \mid B} = \ln(Z)$, with $Z$ the minimum of the sizes of the many-body Hilbert spaces of $A$ and $B$.

A Hamiltonian which acts on a $K$-dimensional quantum lattice system in the thermodynamic limit is called local if there exists a distance cutoff beyond which the interaction terms decay at least exponentially.
Consider the \textit{ground state} $\ket{\Psi_0}$ of a gapped $K$-dimensional quantum system in the thermodynamic limit, and select as subsystem a hypercube with side $L$ and volume $L^K$. The von Neumann entropy is believed to obey an area law \cite{PhysRevLett.94.060503, RevModPhys.82.277, PhysRevLett.111.170501}:
\begin{equation}
S_{\text{hypercube}} \propto L^{K-1}.
\end{equation}
This is the result of a finite correlation length, as only lattice sites in the immediate vicinity of the hypercube's boundary are then correlated with lattice sites on the other side of the boundary. This is a theorem for one-dimensional systems \cite{1742-5468-2007-08-P08024} and a conjecture in higher dimensions \cite{RevModPhys.82.277}, supported by numerical examples and theoretical arguments \cite{PhysRevLett.111.170501}. For critical quantum systems, with a closed excitation gap, there can be logarithmic corrections to the area law \cite{PhysRevLett.90.227902, RevModPhys.82.277}.

For gapped one-dimensional systems, consider as subsystem a line segment of length $L$. Its boundary consists of two points. Due to the finite correlation length in the ground state, the entanglement of the subsystem does not increase with $L$, if $L$ is significantly larger than the correlation length. The von Neumann entropy is then a constant independent of $L$, and the ground state $\ket{\Psi_0}$ can be well represented by retaining only a finite number of states $D$ in the Schmidt decomposition of any bipartition of the lattice in two semi-infinite line segments. This is the reason why the MPS ansatz and the corresponding DMRG algorithm work very well to study the ground states of gapped one-dimensional systems.

\begin{figure}[t!]
\centering
\includegraphics[width=0.45\textwidth]{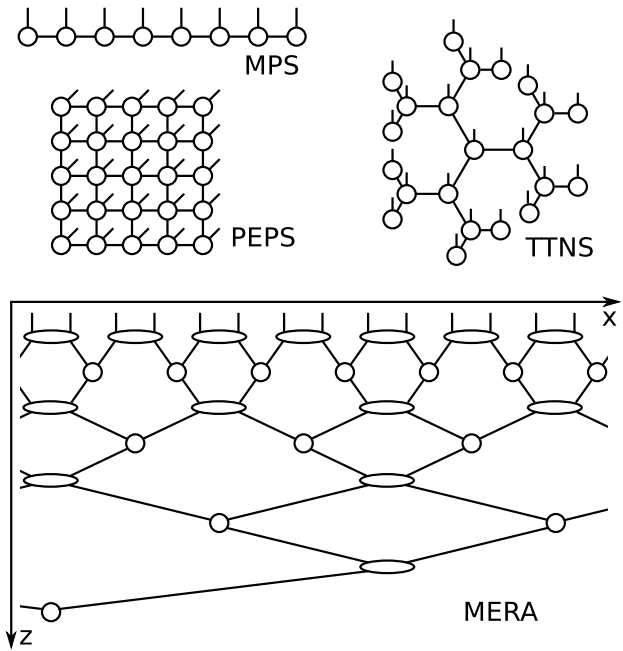}
\caption{\label{TNS-plot} Several tensor network states. Tensors are represented by circles, physical indices by open lines, and virtual indices by connected lines. The graph hence represents how the ansatz decomposes the FCI tensor.}
\end{figure}
The MPS ansatz
\begin{eqnarray}
\ket{\Psi} & = & \sum_{\{n_{j\sigma} \} \{ \alpha_k \}} A[1]^{n_{1\uparrow}n_{1\downarrow}}_{\alpha_1} A[2]^{n_{2\uparrow}n_{2\downarrow}}_{\alpha_1 ; \alpha_2} ... A[L]^{n_{L\uparrow}n_{L\downarrow}}_{\alpha_{L-1}} \nonumber \\ & & \ket{n_{1\uparrow} n_{1\downarrow} n_{2\uparrow} n_{2\downarrow} ... n_{L\uparrow}n_{L\downarrow}}, \label{MPSansatz}
\end{eqnarray}
is shown graphically in Fig. \ref{TNS-plot}. Except for the first and last orbital (or site), the MPS ansatz introduces a rank-3 tensor per site. One of its indices corresponds to the physical index $n_{i\uparrow}n_{i\downarrow}$, the other two to the virtual indices $\alpha_{i-1}$ and $\alpha_i$. Similar to Fig. \ref{FCItoMPS-plot}, tensors are represented by circles, physical indices by open lines, and virtual indices by connected lines in Fig. \ref{TNS-plot}. The graph hence represents how the ansatz decomposes the FCI tensor. The finite size $D$ of the virtual indices can capture finite-length correlations along the one-dimensional chain. Stated more rigorously: for a system in the thermodynamic limit, all correlation functions $C_{\text{MPS}}(\Delta x)$ measured in an MPS ansatz with finite $D$ decay exponentially with increasing site distance $\Delta x$ \cite{FCSmath, TNSoverview}:
\begin{equation}
C_{\text{MPS}}(\Delta x) \propto e^{-\alpha \Delta x}. \label{MPSexpodecaycorrfuncchap2}
\end{equation}

Unless the lattice size is reasonably small \cite{2D-DMRG-citation}, an MPS is not efficient to represent the ground state of higher dimensional or critical systems. Fortunately, efficient tensor network states (TNS) for higher dimensional and critical lattice systems, which do obey the correct entanglement scaling laws, have been developed \cite{TNSoverview}. There even exists a continuous MPS ansatz for one-dimensional quantum field theories \cite{PhysRevLett.104.190405}.

The ansatz for two-dimensional systems is called the projected entangled pair state (PEPS) \cite{PEPS-arxiv}, see Fig. \ref{TNS-plot}. Instead of two virtual indices, each tensor now has four virtual indices,  which allows to arrange the sites in a square lattice. A finite virtual dimension $D$ still introduces a finite correlation length, but due to the topology of the PEPS, this is sufficient for two-dimensional systems, even in the thermodynamic limit. Analogous extensions exist for other lattice topologies.

The ansatz for critical one-dimensional systems is called the multi-scale entanglement renormalization ansatz (MERA) \cite{PhysRevLett.99.220405}, see Fig. \ref{TNS-plot}. This ansatz has two axes: $x$ along the physical one-dimensional lattice and $z$ along the \textit{renormalization direction}. Consider two sites separated by $\Delta x$ along $x$. The number of virtual bonds between those sites is only of order $\Delta z \propto \ln \Delta x$. With finite $D$, all correlation functions $C_{\text{MERA}}(\Delta x)$ measured in a MERA decay exponentially with increasing renormalization distance $\Delta z$:
\begin{equation}
 C_{\text{MERA}}(\Delta x) \propto e^{-\alpha \Delta z} \propto e^{- \beta \ln \Delta x} = (\Delta x)^{-\beta},
\end{equation}
and therefore only algebraically with increasing lattice distance $\Delta x$ \cite{PhysRevLett.99.220405, TNSoverview}.

An inconvenient property of the PEPS, MERA, and MPS with periodic boundary conditions \cite{PhysRevLett.93.227205}, is the introduction of loops in the network. This results in the inability to exploit the TNS gauge invariance to work with orthonormal renormalized environment states, see sections \ref{subsec-can-form-2.3.2} and \ref{subsec-micro-iteraions}. One particular network which avoids such loops, but which is still able to capture polynomially decaying correlation functions, is the tree TNS (TTNS) \cite{TTNS, PhysRevB.87.125139}, see Fig. \ref{TNS-plot}. From a central tensor with $z$ virtual bonds, $Y$ consecutive onion-like layers are built of tensors with also $z$ virtual bonds. The last layer consists of tensors with only 1 virtual bond. An MPS is hence a TTNS with $z=2$. The number of sites $L$ increases as \cite{PhysRevB.82.205105, LegezaTTNS}:
\begin{equation}
 L = 1 + z \sum\limits_{k=1}^{Y} (z-1)^{k-1} = \frac{z (z-1)^{Y} - 2}{z-2}.
\end{equation}
Hence $Y \propto \ln(L)$ for $z \geq 3$. The maximum number of virtual bonds between any two sites is $2Y$. The correlation functions in a TTNS with finite $D$ and $z \geq 3$ decrease exponentially with increasing separation $Y$:
\begin{equation}
 C_{\text{TTNS}}(L) \propto e^{-\alpha Y} \propto e^{- \beta \ln L} = L^{-\beta},
\end{equation}
and therefore only algebraically with increasing number of sites $L$ \cite{TTNS, PhysRevB.87.125139}.

For higher-dimensional or critical systems, DMRG can still be useful \cite{2D-DMRG-citation}. The virtual dimension $D$ then has to be increased to a rather large size to obtain numerical convergence. In the case of multiple dimensions, the question arises if one should work in real or momentum space, and how the corresponding single-particle degrees of freedom should be mapped to the one-dimensional lattice \cite{PhysRevB.53.R10445}. \textit{Ab initio} quantum chemistry can be considered as a higher-dimensional system, due to the full-rank two-body interaction in the Hamiltonian \eqref{QC-ham-2}, and the often compact spatial extent of molecules. Nevertheless, DMRG turned out to be very useful for \textit{ab initio} quantum chemistry (QC-DMRG) \cite{WhiteQCDMRG, QUA:QUA1, mitrushenkov:6815, Chan2002, PhysRevB.67.125114, chan:8551, DMRG_LiF, mitrushenkov:4148, PhysRevB.68.195116, chan:3172, chan:6110, PhysRevB.70.205118, moritz:024107, chan:204101, moritz:184105, moritz:034103, hachmann:144101, Rissler2006519, moritz:244109, dorando:084109, hachmann:134309, marti:014104, zgid:014107, zgid:144115, zgid:144116, ghosh:144117, ChanB805292C, ChanQUA:QUA22099, dorando:184111,kurashige:234114,yanai:024105, neuscamman:024106, RichardsonControlReiher, PhysRevB.81.235129, mizukami:091101, 1367-2630-12-10-103008, PhysRevB.82.205105, Marti:C0CP01883J, PhysRevA.83.012508, boguslawski:224101, kurashige:094104, QUA:QUA23173, sharma:124121, woutersJCP1, JCTCspindens, C2CP23767A, JPCLentanglement, JCTCgrapheneNano, nakatani:134113, JCTCbondForm, naturechem, ma:224105, saitow:044118, Spiropyran, C3CP53975J, NaokiLRTpaper, Knecht-4c-DMRG, 2013arXiv1312.2415W, Harris2014, 2014arXiv1401.5437M, newKura, LegezaTTNS, 1.4867383, C3CP55225J, 10.1063.1.4885815, LegezaLiBe, ChanRevFrontiers, Chan2009149, MartiReiherOldenburg2, chan:annurevphys, WCMS:WCMS1095, KurashigeMolPhys, Accuracy-Keller}. 

An excellent description of QC-DMRG in terms of renormalization transformations is given in Chan and Head-Gordon \cite{Chan2002}. Section \ref{DMRG-sec-chapt-2} contains a description in terms of the underlying MPS ansatz, because this approach will be used in section \ref{SymmetrySection} to introduce $\mathsf{SU(2)} \otimes \mathsf{U(1)} \otimes \mathsf{P}$ symmetry in the DMRG algorithm. The properties of the DMRG algorithm are discussed in section \ref{DMRG-prop}. Several convergence strategies are listed in section \ref{conv_strat_sec}. An overview of the strategies to choose and order orbitals is given in section \ref{DMRG-orb}. A converged DMRG calculation can be the starting point of other methods. These methods are summarized in section \ref{DMRG-algos}. Section \ref{DMRG-systems} gives an overview of the currently existing QC-DMRG codes, and the systems which have been studied with them.

\section{The QC-DMRG algorithm} \label{DMRG-sec-chapt-2}
\subsection{The MPS ansatz} \label{TheMPSansatzSectionInChapterTwo}
DMRG can be formulated as the variational optimization of an MPS ansatz \cite{PhysRevLett.75.3537, PhysRevB.55.2164}. The MPS ansatz \eqref{MPSansatz} has open boundary conditions, because sites 1 and L only have one virtual index. The sites are assumed to be orbitals, which have 4 possible occupancies $\ket{-}$, $\ket{\uparrow}$, $\ket{\downarrow}$, and $\ket{\uparrow\downarrow}$. Henceforth $\ket{n_i}$ will be used as a shorthand for $\ket{n_{i\uparrow}n_{i\downarrow}}$. To be of practical use, the virtual dimensions $\alpha_j$ are truncated to $D$: $\text{dim}(\alpha_j) = \min(4^j, 4^{L-j}, D)$. With increasing $D$, the MPS ansatz spans a larger region of the full Hilbert space, but it is of course not useful to make $D$ larger than $4^{\lfloor \frac{L}{2} \rfloor}$ as the MPS ansatz then spans the whole Hilbert space.

A Slater determinant has gauge freedom: a rotation in the occupied orbital space alone, or a rotation in the virtual orbital space alone, does not change the physical wavefunction. Only occupied-virtual rotations change the wavefunction. An MPS has gauge freedom as well. If for two neighbouring sites $i$ and $i+1$, the left MPS tensors are right-multiplied with the non-singular matrix $G$:
\begin{equation}
\tilde{A}[i]^{n_i}_{\alpha_{i-1};\alpha_i} = \sum\limits_{\beta_i} A[i]^{n_i}_{\alpha_{i-1};\beta_i} G_{\beta_i;\alpha_i},
\end{equation}
and the right MPS tensors are left-multiplied with the inverse of $G$:
\begin{equation}
\tilde{A}[i+1]^{n_{i+1}}_{\alpha_{i};\alpha_{i+1}} = \sum\limits_{\beta_i} G^{-1}_{\alpha_{i};\beta_i} A[i+1]^{n_{i+1}}_{\beta_i;\alpha_{i+1}},
\end{equation}
the wavefunction does not change, i.e. $\forall n_i,n_{i+1},\alpha_{i-1},\alpha_{i+1}$:
\begin{equation}
\sum\limits_{\alpha_i} \tilde{A}[i]^{n_i}_{\alpha_{i-1};\alpha_i} \tilde{A}[i+1]^{n_{i+1}}_{\alpha_{i};\alpha_{i+1}} = \sum\limits_{\alpha_i} A[i]^{n_i}_{\alpha_{i-1};\alpha_i} A[i+1]^{n_{i+1}}_{\alpha_{i};\alpha_{i+1}}.
\end{equation}

\subsection{Canonical forms} \label{subsec-can-form-2.3.2}
The two-site DMRG algorithm consists of consecutive \textit{sweeps} or \textit{macro-iterations}, where at each sweep step the MPS tensors of two neighbouring sites are optimized in the \textit{micro-iteration}. Suppose these sites are $i$ and $i+1$. The gauge freedom of the MPS is used to bring it in a particular canonical form. For all sites to the left of $i$, the MPS tensors are left-normalized:
\begin{equation}
\sum\limits_{\alpha_{k-1}, n_k} \left(A[k]^{n_k}\right)^{\dagger}_{\alpha_k; \alpha_{k-1}} A[k]^{n_k}_{\alpha_{k-1};\beta_k} = \delta_{\alpha_k, \beta_k}, \label{left-normalized}
\end{equation}
and for all sites to the right of $i+1$, the MPS tensors are right-normalized:
\begin{equation}
\sum\limits_{\alpha_{k}, n_k} A[k]^{n_k}_{\alpha_{k-1};\alpha_k} \left(A[k]^{n_k}\right)^{\dagger}_{\alpha_k; \beta_{k-1}} = \delta_{\alpha_{k-1}, \beta_{k-1}}. \label{right-normalized}
\end{equation}
Left-normalization can be performed with consecutive QR-decompositions:
\begin{eqnarray}
& A[k]^{n_k}_{\alpha_{k-1};\alpha_k} = A[k]_{(\alpha_{k-1} n_k) ; \alpha_k} = \nonumber \\
& \sum\limits_{\beta_k} Q[k]_{( \alpha_{k-1} n_k ) ; \beta_k} R_{\beta_k ; \alpha_k} = \sum\limits_{\beta_k} Q[k]^{n_k}_{\alpha_{k-1} ; \beta_k} R_{\beta_k ; \alpha_k}.
\end{eqnarray}
The MPS tensor $Q[k]$ is now left-normalized. The $R$-matrix is multiplied into $A[k+1]$. From site 1 to $i-1$, the MPS tensors are left-normalized this way, without changing the wavefunction. Right-normalization occurs analogously with LQ-decompositions. In section \ref{subsec-macro-it}, it will become clear that this normalization procedure only needs to occur at the start of the DMRG algorithm.

At this point, it is instructive to make the analogy to the renormalization group formulation of the DMRG algorithm. Define the following vectors:
\begin{eqnarray}
\ket{\alpha_{i-1}^L} & = & \sum_{\{n_{j} \} \{ \alpha_1 ... \alpha_{i-2} \}} A[1]^{n_{1}}_{\alpha_1} ... A[i-1]^{n_{i-1}}_{\alpha_{i-2} ; \alpha_{i-1}} \nonumber \\
& & \qquad \qquad \qquad \ket{n_1 ... n_{i-1}},\\
\ket{\alpha_{i+1}^R} & = & \sum_{\{n_{j} \} \{ \alpha_{i+2} ... \alpha_{L-1} \}} A[i+2]^{n_{i+2}}_{\alpha_{i+1} ; \alpha_{i+2}} ... A[L]^{n_{L}}_{\alpha_{L-1}} \nonumber \\
& & \qquad \qquad \qquad \ket{n_{i+2} ... n_{L}}.
\end{eqnarray}
Due to the left- and right-normalization described above, these vectors are orthonormal:
\begin{eqnarray}
\braket{\alpha_{i-1}^L \mid \beta_{i-1}^L } & = & \delta_{\alpha_{i-1}, \beta_{i-1}}, \\
\braket{\alpha_{i+1}^R \mid \beta_{i+1}^R } & = & \delta_{\alpha_{i+1}, \beta_{i+1}}.
\end{eqnarray}
$\{\ket{\alpha_{i-1}^L}\}$ and $\{\ket{\alpha_{i+1}^R}\}$ are renormalized bases of the many-body Hilbert spaces spanned by resp.\ orbitals 1 to $i-1$ and orbitals $i+2$ to $L$. Consider for example the left side. For site $k$ from 1 to $i-2$, the many-body basis is augmented by one orbital and subsequently truncated again to at most $D$ renormalized basis states:
\begin{eqnarray}
& \{ \ket{\alpha_{k-1}^L} \} \otimes \{ \ket{n_k} \} \rightarrow \nonumber \\
& \ket{\alpha_{k}^L} = \sum\limits_{\alpha_{k-1},n_k} A[k]^{n_k}_{\alpha_{k-1};\alpha_k}\ket{\alpha_{k-1}^L} \ket{n_k}.
\end{eqnarray}
DMRG is hence a renormalization group for increasing many-body Hilbert spaces. The next section addresses how this renormalization transformation is chosen.

\subsection{Micro-iterations} \label{subsec-micro-iteraions}
\begin{figure}[t!]
\centering
\includegraphics[width=0.45\textwidth]{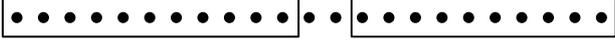}
\caption{\label{Sweeps-plot} Optimization of the MPS tensors at sites $i$ and $i+1$ in the two-site DMRG algorithm. The effective Hamiltonian equation \eqref{effHameq}, obtained by variation of the Lagrangian \eqref{Lagrangian_eq}, can be interpreted as the approximate diagonalization of the exact Hamiltonian $\hat{H}$ in the orthonormal basis $\{ \ket{ \alpha_{i-1}^L } \} \otimes \{ \ket{ n_i } \} \otimes \{ \ket{ n_{i+1} } \} \otimes \{ \ket{ \alpha_{i+1}^R } \}$.}
\end{figure}

Combine the MPS tensors of the two sites under consideration into a single two-site tensor:
\begin{equation}
\sum\limits_{\alpha_i} A[i]^{n_i}_{\alpha_{i-1};\alpha_i} A[i+1]^{n_{i+1}}_{\alpha_{i};\alpha_{i+1}} = B[i]_{\alpha_{i-1};\alpha_{i+1}}^{n_i;n_{i+1}}. \label{two-site-object-not-reduced}
\end{equation}
At the current micro-iteration of the DMRG algorithm, $\mathbf{B}[i]$ (the flattened column form of the tensor $B[i]$) is used as an initial guess for the effective Hamiltonian equation. This equation is obtained by variation of the Lagrangian \cite{ChanB805292C}
\begin{equation}
\mathcal{L} = \braket{\Psi(\mathbf{B}[i]) \mid \hat{H} \mid \Psi(\mathbf{B}[i])} - E_i \braket{\Psi(\mathbf{B}[i]) \mid \Psi(\mathbf{B}[i])} \label{Lagrangian_eq}
\end{equation}
with respect to the complex conjugate of $\mathbf{B}[i]$:
\begin{equation}
\mathbf{H}[i]^{\text{eff}} \mathbf{B}[i] = E_i \mathbf{B}[i]. \label{effHameq}
\end{equation}
The canonical form in Eqs. (\ref{left-normalized})-(\ref{right-normalized}) ensured that no overlap matrix is present in this effective Hamiltonian equation. In the DMRG language, this equation can be interpreted as the approximate diagonalization of the exact Hamiltonian $\hat{H}$ in the orthonormal basis $\{ \ket{ \alpha_{i-1}^L } \} \otimes \{ \ket{ n_i } \} \otimes \{ \ket{ n_{i+1} } \} \otimes \{ \ket{ \alpha_{i+1}^R } \}$, see Fig. \ref{Sweeps-plot}. Because of the underlying MPS ansatz, DMRG is variational: $E_i$ is always an upper bound to the energy of the true ground state.

The lowest eigenvalue and corresponding eigenvector of the effective Hamiltonian are searched with iterative sparse eigensolvers. Typical choices are the Lanczos or Davidson algorithms \cite{Lanczos, Davidson197587}. Once $\mathbf{B}[i]$ is found, it is decomposed with an SVD:
\begin{eqnarray}
B[i]_{ \left( \alpha_{i-1} n_i \right) ; \left( n_{i+1} \alpha_{i+1} \right)} = \nonumber \\
\sum\limits_{\beta_i} U[i]_{ \left( \alpha_{i-1} n_i \right) ; \beta_i} \kappa[i]_{\beta_i} V[i]_{ \beta_i ; \left( n_{i+1} \alpha_{i+1} \right)} . \label{SVDofBsolution}
\end{eqnarray}
Note that $U[i]$ is hence left-normalized and $V[i]$ right-normalized. The sum over $\beta_i$ is truncated if there are more than $D$ nonzero Schmidt values $\kappa[i]_{\beta_i}$, thereby keeping the $D$ largest ones. This is the optimal approximation for the bipartition of $\{ \ket{ \alpha_{i-1}^L } \} \otimes \{ \ket{ n_i } \} \otimes \{ \ket{ n_{i+1} } \} \otimes \{ \ket{ \alpha_{i+1}^R } \}$ into $A = \{ \ket{ \alpha_{i-1}^L } \} \otimes \{ \ket{n_i} \}$ and $B = \{ \ket{n_{i+1}} \} \otimes \ket{ \alpha_{i+1}^R } \}$. In the original DMRG algorithm, $U[i]$ and $V[i]$ were obtained as the eigenvectors of resp.\ $\hat{\rho}^A$ and $\hat{\rho}^B$.

A discarded weight can be associated with the truncation of the sum over $\beta_i$:
\begin{equation}
 w[i]^{\text{disc}}_D = \sum\limits_{\beta_i > D} \kappa[i]^2_{\beta_i}.
\end{equation}
This is the probability to measure one of the discarded states in the subsystems $A$ or $B$. The approximation introduced by the truncation becomes better with increasingly small discarded weight. Instead of working with a fixed $D$, one could also choose $D$ dynamically in order to keep $w[i]^{\text{disc}}_D$ below a preset threshold, as is done in Legeza's dynamic block state selection approach \cite{PhysRevB.67.125114}.

\subsection{Macro-iterations or sweeps} \label{subsec-macro-it}

So far, we have looked at a micro-iteration of the DMRG algorithm. This micro-iteration happens during left or right sweeps. During a left sweep, $B[i]$ is constructed, the corresponding effective Hamiltonian equation solved, the solution $B[i]$ decomposed, the Schmidt spectrum truncated, $\kappa[i]$ is contracted into $U[i]$, $A[i]$ is set to this contraction $U[i] \times \kappa[i]$, $A[i+1]$ is set to $V[i]$, and $i$ is decreased by 1. Note that $A[i+1]$ is right-normalized for the next micro-iteration as required. This stepping to the left occurs until $i=1$, and then the sweep direction is reversed from left to right. Based on energy differences, or wavefunction overlaps, between consecutive sweeps, a convergence criterium is triggered, and the sweeping stops.

DMRG can be regarded as a self-consistent field method: at convergence the neighbours of an MPS tensor generate the field which yields the local solution, and this local solution generates the field for its neighbours \cite{Chan2002, hachmann:144101, ChanB805292C}.

\subsection{Renormalized operators and their complements}
The effective Hamiltonian in Eq. (\ref{effHameq}) is too large to be fully constructed as a matrix. Only its action on a particular guess $\mathbf{B}[i]$ is available as a function. In order to construct $\mathbf{H}[i]^{\text{eff}} \mathbf{B}[i]$ efficiently for general quantum chemistry Hamiltonians, several tricks are used. Suppose that a right sweep is performed and that the MPS tensors of sites $i$ and $i+1$ are about to be optimized.

Renormalized operators such as $\braket{\alpha_{i-1}^L \mid \hat{a}_{k\sigma}^{\dagger} \hat{a}_{l\tau} \mid \beta_{i-1}^L}$ with $k,l \leq i-1$ are constructed and stored on disk \cite{WhiteQCDMRG, Chan2002, kurashige:234114}. The renormalized operators needed for the previous micro-iteration can be recycled to this end. Suppose $k,l \leq i-2$:
\begin{eqnarray}
& \braket{\alpha_{i-1}^L \mid \hat{a}_{k\sigma}^{\dagger} \hat{a}_{l\tau} \mid \beta_{i-1}^L} = \sum\limits_{\alpha_{i-2}, \beta_{i-2}, n_{i-1}} \left(A[i-1]^{n_{i-1}}\right)^{\dagger}_{\alpha_{i-1} ; \alpha_{i-2}} \nonumber \\
& \braket{\alpha_{i-2}^L \mid \hat{a}_{k\sigma}^{\dagger} \hat{a}_{l\tau} \mid \beta_{i-2}^L} A[i-1]^{n_{i-1}}_{\beta_{i-2} ; \beta_{i-1}}. \label{operator_tfo}
\end{eqnarray}
Note that no phases appear because an even number of second-quantized operators was transformed. For an odd number, there should be an additional phase $(-1)^{n_{(i-1)\uparrow} + n_{(i-1)\downarrow}}$ at the right-hand side (RHS) due to the Jordan-Wigner transformation \cite{JordanWignerTfo}. Renormalized operators to the right of $B[i]$ can be loaded from disk, as they have been saved during the previous left sweep.

Once three second-quantized operators are on one side of $B[i]$, they are multiplied with the matrix elements $h_{kl;mn}$, and a summation is performed over the common indices to construct \textit{complementary} renormalized operators \cite{PhysRevB.53.R10445, WhiteQCDMRG, Chan2002, kurashige:234114}:
\begin{eqnarray}
\braket{\alpha_{i-1}^L \mid \hat{Q}_{n\tau} \mid \beta_{i-1}^L} = \nonumber \\
\sum\limits_{\sigma} \sum\limits_{k,l,m<i} h_{kl;mn} \braket{\alpha_{i-1}^L \mid \hat{a}_{k\sigma}^{\dagger} \hat{a}^{\dagger}_{l\tau} \hat{a}_{m\sigma} \mid \beta_{i-1}^L}. \label{Q_complement_op}
\end{eqnarray}
For two, three, and four second-quantized operators on one side of $B[i]$, these complementary renormalized operators are constructed. A bare renormalized operator (without matrix elements) is only constructed for one or two second-quantized operators.

Hermitian conjugation and commutation relations:
\begin{eqnarray}
& \braket{\alpha_{i-1}^L \mid \hat{a}_{k\sigma}^{\dagger} \hat{a}^{\dagger}_{l\tau} \mid \beta_{i-1}^L} = \braket{\beta_{i-1}^L \mid \hat{a}_{l\tau} \hat{a}_{k\sigma} \mid \alpha_{i-1}^L}^{\dagger} \nonumber \\
& = - \braket{\alpha_{i-1}^L \mid \hat{a}^{\dagger}_{l\tau} \hat{a}_{k\sigma}^{\dagger} \mid \beta_{i-1}^L},
\end{eqnarray}
are also used to further limit the storage requirements for the (complementary) renormalized operators. Examples of renormalized operators and the fermion sign handling can be found in, for example, Refs. \cite{woutersJCP1, my_phd}.

\subsection{Computational cost}
This section describes the cost of the QC-DMRG algorithm per sweep in terms of memory, disk, and computational time \cite{WhiteQCDMRG, Chan2002, kurashige:234114}. To analyze this cost, let us first look at the cost per micro-iteration. A micro-iteration consists of three steps: solving the effective Hamiltonian equation \eqref{effHameq}, performing an SVD of the solution \eqref{SVDofBsolution}, and constructing the (complementary) renormalized operators for the next micro-iteration.

To solve the effective Hamiltonian equation with the Lanczos or Davidson algorithms, a set of $N_{vec}$ trial vectors $\{ \mathbf{B}[i] \}$ are kept in memory, as well as $\mathbf{H}[i]^{\text{eff}} \{ \mathbf{B}[i] \}$. To construct $\mathbf{H}[i]^{\text{eff}} \{ \mathbf{B}[i] \}$, (complementary) renormalized operators should also be stored in memory. The latter have at most two site indices. The total memory cost is hence $\mathcal{O}((N_{vec} + L^2) D^2)$.

The action of $\mathbf{H}[i]^{\text{eff}}$ on $\mathbf{B}[i]$ is divided into several contributions. Each contribution consists of the joint action of a renormalized operator and the corresponding complementary renormalized operator. For each contribution, two matrix-matrix multiplications need to be performed, of computational cost $\mathcal{O}(D^3)$. In total there are $\mathcal{O}(L^2)$ contributions, because complementary renormalized operators have at most two site indices. The total computational cost is hence $\mathcal{O}(N_{\text{vec}} L^2 D^3)$ for the multiplications, and $\mathcal{O}(N_{\text{vec}} L^2 D^2)$ for the summation of the different contributions.

The SVD of the solution $\mathbf{B}[i]$ takes $\mathcal{O}(D^3)$ computational time and $\mathcal{O}(D^2)$ memory.

The construction of one particular renormalized operator takes $\mathcal{O}(D^3)$ computational time and $\mathcal{O}(D^2)$ memory, and there are $\mathcal{O}(L^2)$ such operators. The most tedious part to analyze is the construction of the two-site complementary renormalized operators, e.g.
\begin{equation}
\braket{\alpha_{i-1}^L \mid \hat{F}_{m \sigma ; n\tau} \mid \beta_{i-1}^L} = \sum\limits_{k,l<i} h_{kl;mn} \braket{\alpha_{i-1}^L \mid \hat{a}_{k\sigma}^{\dagger} \hat{a}^{\dagger}_{l\tau} \mid \beta_{i-1}^L},
\end{equation}
which takes at first sight $\mathcal{O}(L^2 D^2)$ computational time and $\mathcal{O}(D^2)$ memory per operator. There are $\mathcal{O}(L^2)$ such operators, and a naive implementation would hence result in a computational cost of $\mathcal{O}(L^4 D^2)$ per micro-iteration. However, this summation needs to be performed only once for each operator, at the moment when the second second-quantized operator is added:
\begin{eqnarray}
\braket{\alpha_{i-1}^L \mid \hat{F}_{m \sigma ; n\tau} \mid \beta_{i-1}^L} = \nonumber \\
\sum\limits_{k<i} h_{k(i-1);mn} \braket{\alpha_{i-1}^L \mid \hat{a}_{k\sigma}^{\dagger} \hat{a}^{\dagger}_{(i-1)\tau} \mid \beta_{i-1}^L}. \label{smart-two-index-complements}
\end{eqnarray}
From then on, this operator can be transformed as in Eq. \eqref{operator_tfo}. The total computational cost per micro-iteration is hence reduced to $\mathcal{O}(L^3 D^2)$ for the summation (there are three variable site indices in Eq. \eqref{smart-two-index-complements}), and $\mathcal{O}(L^2 D^3)$ for the transformation (there are $\mathcal{O}(L^2)$ operators to be transformed). The one-site complementary renormalized operator (the complement of three second-quantized operators) can be constructed from the two-site complementary renormalized operators at the moment when the third second-quantized operator is added. From then on, this operator can also be transformed as in Eq. \eqref{operator_tfo}.

\begin{figure}[t!]
\centering
\includegraphics[width=0.48\textwidth]{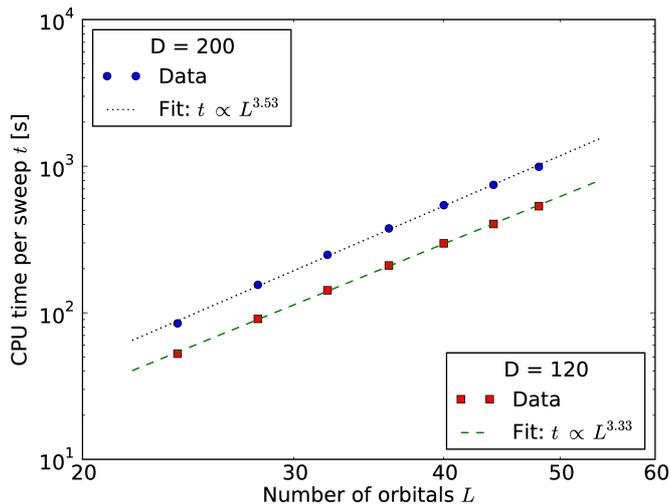}
\caption{\label{Timings-plot} The geometries of all-trans polyenes C$_n$H$_{n+2}$ were optimized at the B3LYP/6-31G** level of theory for $n=12$, 14, 16, 18, 20, 22 and 24. The $\sigma$-orbitals were kept frozen at the RHF/6-31G level of theory. The $\pi$-orbitals in the 6-31G basis were localized by means of the Edmiston-Ruedenberg localization procedure \cite{RevModPhys.35.457}, which maximizes $\sum_i v_{ii;ii}$. The localized $\pi$-orbitals belong to the $A''$ irrep of the $\mathsf{C_s}$ point group, and were ordered according to the one-dimensional topology of the polyene. For all polyenes, the average CPU time per DMRG sweep was determined with \textsc{CheMPS2} \cite{2013arXiv1312.2415W}, for two \textit{reduced} virtual dimensions $D$. For the values of $D$ shown here, the energies are converged to $\mu E_h$ accuracy due to the one-dimensional topology of the localized and ordered $\pi$-orbitals. Due to the imposed $\mathsf{SU(2)} \otimes \mathsf{U(1)} \otimes \mathsf{C_s}$ symmetry, all tensors become block-sparse, see section \ref{SymmetrySection}, which causes the scaling to be below $\mathcal{O}(L^4)$.}
\end{figure}

As mentioned earlier, the (complementary) renormalized operators are stored to disk, as well as the MPS site tensors, in order to be recycled when the sweep direction is reversed. An overview of the resulting total cost per \textit{macro-iteration} is given in Tab. \ref{compuReqDMRG}. For a given virtual dimension $D$, the DMRG algorithm is of polynomial cost in $L$. The computational requirements in Tab. \ref{compuReqDMRG} are upper bounds if the symmetry group of the Hamiltonian is exploited, see section \ref{SymmetrySection}. Then the MPS tensors and corresponding (complementary) renormalized operators become block-sparse, and $h_{kl;mn}$ is not full rank. An example of the scaling of the computational time per DMRG sweep with the number of orbitals $L$ is shown in Fig. \ref{Timings-plot}. Due to the imposed $\mathsf{SU(2)} \otimes \mathsf{U(1)} \otimes \mathsf{C_s}$ symmetry, \textsc{CheMPS2} \cite{2013arXiv1312.2415W} achieves a scaling below $\mathcal{O}(L^4)$.

\begin{table*}[ht!]
\centering
\caption{\label{compuReqDMRG} Computational requirements per \textit{macro-iteration} or \textit{sweep} of the QC-DMRG algorithm.}
\begin{tabular}{|l|rrr|}
\hline
$\mathcal{O}(\text{task})$ & time & memory & disk \\
\hline
$\mathbf{H}[i]^{\text{eff}} \{ \mathbf{B}[i] \}~^{(a)}$  & $N_{vec} L^3 D^3$    & $N_{vec} D^2$ & - \\
SVD and basis truncation                          & $L D^3$                & $D^2$              & $L D^2$ \\
Renormalized operators                            & $L^3 D^3$            & $L^2 D^2$            & $L^3 D^2$ \\
Complementary renormalized operators              & $L^4 D^2 + L^3 D^3$  & $L^2 D^2$            & $L^3 D^2$ \\
\hline
Total              & $L^4 D^2 + N_{vec} L^3 D^3$ & $(N_{\text{vec}} + L^2) D^2$ & $L^3 D^2$ \\
\hline
\end{tabular}

{$^{(a)}$ The memory for the (complementary) renormalized operators is mentioned separately.}
\end{table*}

\section{Properties} \label{DMRG-prop}
\subsection{DMRG is variational}
The DMRG algorithm is variational, because it can be formulated as the optimization of an MPS ansatz. All energies obtained during all micro-iterations are therefore upper bounds to the true ground state energy. These energies do not go down monotonically however, because the basis $\{ \ket{ \alpha_{i-1}^L } \} \otimes \{ \ket{ n_i } \} \otimes \{ \ket{ n_{i+1} } \} \otimes \{ \ket{ \alpha_{i+1}^R } \}$ in which $\hat{H}$ is diagonalized changes between different micro-iterations due to the truncation of the Schmidt spectrum \cite{Chan2002}.

\subsection{Energy extrapolation}
\begin{figure}
\centering
\includegraphics[width=0.48\textwidth]{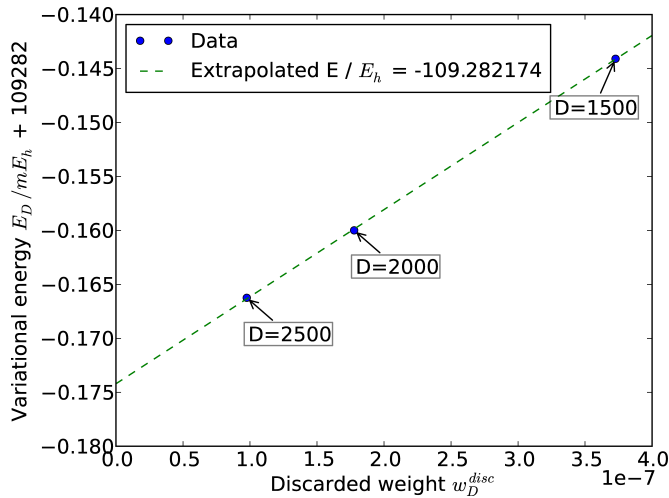}
\caption{\label{ExtrapolN2-plot} Extrapolation of the variational DMRG ground-state energy $E_D$ with the discarded weight $w^{\text{disc}}_D$, for N$_2$ in the cc-pVDZ basis near equilibrium (nuclear separation 2.118 a.u.). The calculation was performed with \textsc{CheMPS2} \cite{2013arXiv1312.2415W} with $\mathsf{SU(2)} \otimes \mathsf{U(1)} \otimes \mathsf{D_{2h}}$ symmetry, see section \ref{SymmetrySection}. $D$ denotes the number of \textit{reduced} virtual basis states. The irrep ordering in the DMRG calculation was [A$_g$B$_{1u}$B$_{3u}$B$_{2g}$B$_{2u}$B$_{3g}$B$_{1g}$A$_{u}$] in order to place bonding and antibonding orbitals close to each other on the one-dimensional DMRG lattice, see section \ref{subsec-entang-meas} and Fig. \ref{ConvergenceN2e-plot}.}
\end{figure}

With increasing virtual dimension $D$, the MPS ansatz spans an increasing part of the many-body Hilbert space. In the following, $E_D$ denotes the minimum energy encountered in Eq. \eqref{effHameq} during the micro-iterations for a given virtual dimension $D$. Several calculations with increasing $D$ can be performed, in order to assess the convergence. This even allows to make an extrapolation of the energy to the FCI limit. Several extrapolation schemes have been suggested. Note that $E_{\text{FCI}}$ and $\{ C_i, p_j, q_k \}$ below are parameters to be fitted. The maximum discarded weight encountered during the last sweep before convergence is abbreviated as:
\begin{equation}
w^{\text{disc}}_D = \max\limits_i \left\{ w[i]^{\text{disc}}_D \right\}.
\end{equation}
The initial assumption of exponential convergence \cite{WhiteQCDMRG}
\begin{equation}
\ln \left( E_D - E_{\text{FCI}} \right) \propto C_1 + C_2 D \label{ExtrapolExpon}
\end{equation}
was rapidly abandoned for the relation \cite{PhysRevB.53.14349,Chan2002,PhysRevB.67.125114}
\begin{equation}
E_D - E_{\text{FCI}} = C_3 w^{\text{disc}}_D, \label{Eextrapol1}
\end{equation}
because the energy is a linear function of the RDM \cite{Chan2002}. An example of an extrapolation with Eq. \eqref{Eextrapol1} is shown in Fig. \ref{ExtrapolN2-plot}. The tail of the distribution of RDM eigenvalues scales as \cite{AyersRDMcalues, Chan2002}
\begin{equation}
\kappa[i]^2_{\beta_i} \propto \exp \left\{ - C_4 \left( \ln \beta_i \right)^2 \right\}. \label{schmidtspectrumdecayrelationcroot}
\end{equation}
Substituting this relation in Eq. \eqref{Eextrapol1} yields an improved version of Eq. \eqref{ExtrapolExpon} \cite{Chan2002}:
\begin{equation}
\ln \left( E_D - E_{\text{FCI}} \right) \propto C_5 - C_4 \left( \ln D \right)^2. \label{Eextrapol2}
\end{equation}
An example of an extrapolation with Eq. \eqref{Eextrapol2} is shown in Fig. \ref{ConvergenceHubbardSymmetry-plot}. Eqs. \eqref{Eextrapol1} and \eqref{Eextrapol2} are the most widely used extrapolation schemes in QC-DMRG. Three other relations have been proposed as well. A relation for incremental energies $\Delta E_{D_1} = E_{D_1} - E_{D_0}$ has been suggested \cite{mitrushenkov:4148}:
\begin{equation}
\Delta E_{D} = \frac{C_6 + C_7 E_D}{\sqrt{L^3D^2 + 2L^2D^3}},
\end{equation}
but the extrapolated $E_{\text{FCI}}$ often violates the variational principle. An alternative relation based on the discarded weight has also been proposed \cite{mitrushenkov:4148}:
\begin{equation}
\ln \left( E_D - E_{\text{FCI}} \right) = C_8 - C_9 \left( w^{\text{disc}}_D \right)^{-\frac{1}{2}},
\end{equation}
as well as a Richardson-type extrapolation scheme, based on the assumption that the energy is an analytic function of $w^{\text{disc}}_D$ \cite{RichardsonControlReiher}:
\begin{equation}
E^{(\mu\nu)}(w^{\text{disc}}_D) = \frac{p_0 + p_1 w^{\text{disc}}_D + ... + p_{\mu} \left( w^{\text{disc}}_D \right)^{\mu}}{q_0 + q_1 w^{\text{disc}}_D + ... + q_{\nu} \left( w^{\text{disc}}_D \right)^{\nu}}.
\end{equation}

\subsection{The CI content of the wavefunction}
To analyze the MPS wavefunction \eqref{MPSansatz}, suppose that the $L$ orthonormal orbitals are the HF orbitals. An important difference with traditional post-HF methods such as CI expansions, is that no FCI coefficients are a priori zero. An MPS hence captures CI coefficients of any particle-excitation rank relative to HF \cite{chan:6110, hachmann:144101}. A small virtual dimension implies little information content in the FCI coefficient tensor, or equivalently that the many nonzero FCI coefficients are in fact highly correlated. This has to be contrasted with CI expansions, which are truncated in their particle-excitation rank and therefore set many FCI coefficients a priori to zero. The nonzero FCI coefficients are however not a priori correlated in a CI expansion: they are entirely free to be variationally optimized.

\subsection{Size-consistency}
For a method to be size-consistent, the compound wavefunction should be multiplicatively separable $\ket{\Psi} = \ket{A} \ket{B}$ and the energy additively separable $E = E_A + E_B$ for noninteracting subsystems $A$ and $B$. From the discussion of the Schmidt decomposition above, it follows immediately that an MPS is size-consistent if the orbitals of subsystems $A$ and $B$ do not overlap, and if they are separated into two groups on the one-dimensional DMRG lattice \cite{Chan2002, chan:annurevphys}. The latter is for example realized if orbitals $1$ to $k$ correspond to subsystem $A$ and orbitals $k+1$ to $L$ correspond to subsystem $B$. DMRG will then automatically retrieve a product wavefunction, in which only one Schmidt value is nonzero at the corresponding boundary.

\subsection{DMRG is not FCI}
A good variational energy does not necessarily imply that the wavefunction is accurate. Suppose we have an orthonormal MPS $\ket{\Psi_{\text{MPS}}}$ with virtual dimension $D$ which has been variationally optimized to approximate the true ground state $\ket{\Psi_{0}}$ with energy $E_0$. Suppose that
\begin{equation}
\ket{\Psi_{\text{MPS}}} = \sqrt{1 - \epsilon^2} \ket{\Psi_{0}} + \epsilon \ket{\widetilde{\Psi}}
\end{equation}
with $\braket{ \Psi_{0} \mid \widetilde{\Psi} } = 0$. Then
\begin{eqnarray}
\| \ket{\Psi_{\text{MPS}}} - \ket{\Psi_{0}} \|_2 & = & \sqrt{\left(\sqrt{1-\epsilon^2} - 1\right)^2 + \epsilon^2} \nonumber \\
& = & \epsilon + \mathcal{O}(\epsilon^3)
\end{eqnarray}
and
\begin{equation}
\braket{\Psi_{\text{MPS}} \mid \hat{H} \mid \Psi_{\text{MPS}}} - E_{0} = \epsilon^2 \left( \braket{\widetilde{\Psi} \mid \hat{H} \mid \widetilde{\Psi}} - E_{0} \right).
\end{equation}
The energy converges quadratically in the wavefunction error. Most DMRG convergence criteria rely on energy convergence ($\epsilon^2 \approx 0$), see Fig. \ref{ExtrapolN2-plot}. An important implication is that, except for tremendously large virtual dimensions $D$ where $\epsilon \approx 0$, the MPS wavefunction is not invariant to orbital rotations. The orbital choice and their ordering on a one-dimensional lattice also influence the convergence rate with $D$. Strategies to choose and order orbitals are discussed in section \ref{DMRG-orb}. Sparse iterative FCI eigensolvers converge the FCI tensor to a predefined threshold instead of the energy. An FCI solution can therefore be considered invariant to orbital rotations.

\section{Convergence strategies} \label{conv_strat_sec}
The DMRG algorithm can get stuck in a local minimum or a limit cycle, if $D$ is insufficiently large \cite{Chan2002}. The chance of occurrence is larger for inconvenient orbital choices and orderings. Because the virtual dimension $D$ cannot be increased indefinitely in practice, it is important to choose the set of orbitals and their ordering well, see section \ref{DMRG-orb}. Additional considerations to enhance convergence are described here.

\subsection{The number of sites to be optimized in a micro-iteration}
It is better to use the two-site DMRG algorithm than the one-site version \cite{PhysRevB.72.180403}. In the one-site version, the Hamiltonian $\hat{H}$ is diagonalized during the micro-iterations in the basis $\{ \ket{ \alpha_{i-1}^L } \} \otimes \{ \ket{ n_i } \} \otimes \{ \ket{ \alpha_{i}^R } \}$ instead of $\{ \ket{ \alpha_{i-1}^L } \} \otimes \{ \ket{ n_i } \} \otimes \{ \ket{ n_{i+1} } \} \otimes \{ \ket{ \alpha_{i+1}^R } \}$. Because of the larger variational freedom in the two-site DMRG algorithm, lower energy solutions are obtained, and the algorithm is less likely to get stuck \cite{zgid:144115}. It might therefore be worthwhile to optimize three or more MPS tensors simultaneously in a micro-iteration, or to group several orbitals into a single DMRG lattice site \cite{WhiteQCDMRG}.

The two-site algorithm has another important advantage, when the symmetry group of the Hamiltonian is exploited. The virtual dimension $D$ is then distributed over several symmetry sectors, see section \ref{SymmetrySection}. In the one-site algorithm, the virtual dimension of a symmetry sector has to be changed manually during the sweeps \cite{zgid:144115}, while the SVD \eqref{SVDofBsolution} in the two-site algorithm automatically picks the best distribution.

\subsection{Perturbative corrections and noise}
White suggested to add perturbative corrections to the RDM in order to enhance convergence \cite{PhysRevB.72.180403}. Instead of using perturbative corrections, one can also add noise to the RDM prior to diagonalization or to $B[i]$ prior to SVD \cite{Chan2002}. The corrections or noise help to reintroduce lost symmetry sectors (lost quantum numbers) in the renormalized basis, which are important for the true ground state. Instead of adding noise or perturbative corrections, one can also reserve a certain percentage of the virtual dimension $D$ to be distributed equally over all symmetry sectors \cite{chan:3172}.

\subsection{Getting started}
The wavefunction from which the QC-DMRG algorithm starts has an influence on the converged energy (by getting stuck in a local minimum) and on the rate of convergence \cite{PhysRevB.67.125114, PhysRevB.68.195116, moritz:034103}. The effect of the starting guess is estimated to be an order of magnitude smaller than the effect of the choice and ordering of the orbitals \cite{moritz:034103}. Nevertheless, it deserves attention.

One possibility is to choose a small active space to start from, and subsequently augment this active space stepwise with previously frozen orbitals \cite{mitrushenkov:6815}, in analogy to the infinite-system DMRG algorithm \cite{PhysRevLett.69.2863}. Natural orbitals from a small CASSCF calculation or HF orbitals can be used to this end \cite{moritz:034103}. An alternative is to make an a priori guess of how correlated the orbitals are. This can be done with a DMRG calculation with small virtual dimension $D$, from which the approximate single-orbital entropies can be obtained. The subsystem $A$ is then chosen to be a single orbital in Eq. \eqref{von-neumann-entropy-eq}. The larger the single-orbital entropy, the more it is correlated. The active space can then be chosen and dynamically extended based on the single-orbital entropies \cite{PhysRevA.83.012508}.

One can also decompose the wavefunction from a cheap CI calculation with single and double excitations into an MPS to start from \cite{Chan2002, moritz:034103}. Another possibility is to distribute $D$ equally over the symmetry sectors, and to fill the MPS with noise. This retrieves energies below the HF energy well within the first macro-iteration \cite{2013arXiv1312.2415W, my_phd}.

To achieve a very accurate MPS quickly, it is also best to start from calculations with relatively small virtual dimension $D$, and to enlarge it stepwise \cite{Chan2002, moritz:034103, PhysRevLett.77.3633}.

\section{Orbital choice and ordering} \label{DMRG-orb}
There are many ways to set up a renormalization group flow, and the specific setup influences the outcome. One consideration of key importance in QC-DMRG is the choice and ordering of orbitals. Most molecules or active spaces are far from one-dimensional. By placing the orbitals on a one-dimensional lattice, and by assuming an MPS ansatz with modest $D$, an artifical correlation length is introduced in the system, which can be a bad approximation. Over time, several rules of thumb have been established to choose and order the orbitals.

\subsection{Elongated molecules}
\begin{figure}
\centering
\includegraphics[width=0.48\textwidth]{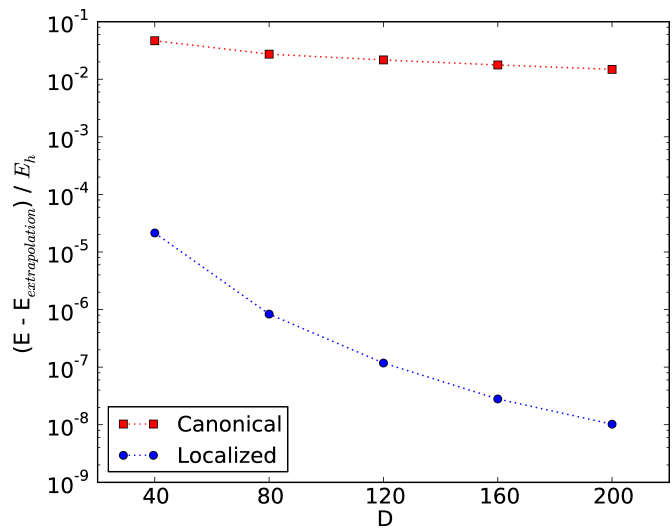} 
\caption{\label{ConvergencePolyene-plot} The computational details were discussed in the caption of Fig. \ref{Timings-plot}. The active space of C$_{14}$H$_{16}$, which consists of 28 $\pi$-orbitals, is studied both with ordered localized orbitals (Edmiston-Ruedenberg) and canonical orbitals (restricted HF). The energy converges significantly faster with the number of \textit{reduced} virtual basis states $D$ when ordered localized orbitals are used.}
\end{figure}

Quantum information theory learns that locality is an important concept, see section \ref{entanglement-section}. The Coulomb interaction is however long-ranged. On the other hand, the mutual screening of electrons and nuclei can result in an effectively local interaction. For elongated molecules such as hydrogen chains \cite{Chan2002, hachmann:144101, QUA:QUA23173, woutersJCP1, nakatani:134113, ma:224105}, polyenes \cite{Chan2002, chan:204101, hachmann:144101, ghosh:144117, yanai:024105}, or acenes \cite{hachmann:134309, dorando:084109, JCTCgrapheneNano}, which are more or less one-dimensional, choosing a spatially local basis has turned out to be very beneficial. There are roughly three ways to choose a local basis: symmetric orthogonalization as it lies closest to the original gaussian basis functions \cite{hachmann:134309, dorando:084109, QUA:QUA23173, woutersJCP1, ma:224105, PhysRev.105.102}, explicit localization procedures such as Pipek-Mezey or Edmiston-Ruedenberg \cite{ghosh:144117, JCTCgrapheneNano, Pipek-Mezey, RevModPhys.35.457}, and working in a biorthogonal basis \cite{chan:204101, QUA:QUA23173}. For the latter, the effective Hamiltonian is not hermitian anymore. The DMRG algorithm should then be correspondingly adapted \cite{Mitru_arxiv, chan:204101, QUA:QUA23173}. The adapted algorithm is slower and prone to convergence issues, and it is therefore better to use one of the other two localized bases \cite{chan:204101, QUA:QUA23173}. Fig. \ref{ConvergencePolyene-plot} illustrates the speed-up in energy convergence by using a localized basis for all-trans polyenes.

\subsection{Hamiltonian measures} \label{sec-integral-measures}
If the topology of the molecule does not provide hints for choosing and ordering orbitals, it was investigated whether the Hamiltonian \eqref{QC-ham} can be of use. Several integral measures have been proposed, for which a minimal bandwidth is believed to yield a good orbital order. Chan and Head-Gordon proposed to minimize the bandwidth of the one-electron integral matrix $t_{ij}$ of the HF orbitals \cite{Chan2002}. In quantum chemistry, it is often stated that the one-electron integrals are an order of magnitude larger than the two-electron integrals, and that quantum chemistry therefore corresponds to the small-$U$ limit of the Hubbard model \cite{PhysRevB.67.125114, PhysRevA.83.012508, Hubbard26111963}. On the other hand, there are many two-electron integrals, and they may become important due to their number. When other orbitals than the HF orbitals are used, it may therefore be interesting to minimize the bandwidth of the Fock matrix \cite{DMRG_LiF}:
\begin{equation}
F_{ij} = t_{ij} + \sum\limits_{k \in \text{occ}} \left( 4 v_{ik;jk} - 2 v_{ik;kj} \right).
\end{equation}
Other proposed integral measures are the MP2-inspired matrix \cite{mitrushenkov:4148}:
\begin{equation}
G_{ij} = \frac{ v_{ii;jj}^2}{| \epsilon_i - \epsilon_j |}
\end{equation}
where $\{ \epsilon_i \}$ are the HF single-particle energies, as well as several measures in Ref. \cite{moritz:024107}. These are the Coulomb matrix $J_{ij} = v_{ij;ij}$, the exchange matrix $K_{ij} = v_{ij;ji}$, the mean-field matrix $M_{ij} = \left( 2 J_{ij} - K_{ij} \right)$, and two derived quantities:
\begin{eqnarray}
J_{ij}^{'} & = & e^{-J_{ij}} \\
M_{ij}^{'} & = & e^{-M_{ij}}.
\end{eqnarray}
While the one-electron integrals $t_{ij}$ vanish when orbitals $i$ and $j$ belong to different molecular point group irreps, $J_{ij}$ and $K_{ij}$ do not. Ref. \cite{moritz:024107} used a genetic algorithm to find the optimal HF orbital ordering, in order to assess the proposed integral measures. This genetic algorithm was expensive, which limited its usage to small test systems. It favoured $K_{ij}$ bandwidth minimization, although no definite conclusions were drawn \cite{moritz:024107}. The exchange matrix $K_{ij}$ was recently used in two DMRG studies \cite{JCTCgrapheneNano,nakatani:134113} in conjunction with localized orbitals, because it then directly reflects their overlaps and distances.

\subsection{Entanglement measures} \label{subsec-entang-meas}
DMRG can be analyzed by means of the underlying MPS ansatz and quantum information theory. The latter can tell us something more than locality. Legeza and S\'olyom proposed to use the single-orbital entropies to find an optimal ordering \cite{PhysRevB.68.195116}. Subsystem A is then chosen to be a single orbital $k$ in Eq. \eqref{von-neumann-entropy-eq}, and its entropy is denoted by $S_1(k)$. It can be efficiently calculated in the DMRG algorithm, because the corresponding RDM $\hat{\rho}^k$ can be built from the expectation values $\braket{(1-\hat{n}_{k\uparrow})(1- \hat{n}_{k\downarrow})}$, $\braket{\hat{n}_{k\uparrow} \hat{n}_{k\downarrow}}$, $\braket{\hat{n}_{k\uparrow} (1-\hat{n}_{k\downarrow})}$, and $\braket{(1-\hat{n}_{k\uparrow}) \hat{n}_{k\downarrow}}$, in which $\hat{n}_{k\sigma} = \hat{a}_{k\sigma}^{\dagger}\hat{a}_{k\sigma}$ \cite{Rissler2006519}. This procedure hence does not require to reorder any orbitals. The larger the single-orbital entropy $S_1(k)$, the more orbital $k$ is correlated. Legeza and S\'olyom proposed to perform a small-$D$ DMRG calculation to estimate $S_1(k)$, and to place the orbitals with large $S_1(k)$ in the center of the chain, and the ones with small $S_1(k)$ near the edges. They reasoned that orbitals close to the Fermi surface are more entangled and therefore have a larger single-orbital entropy. Because DMRG only captures local correlations, these orbitals should lie close to each other.

\begin{figure}
\centering
\includegraphics[width=0.48\textwidth]{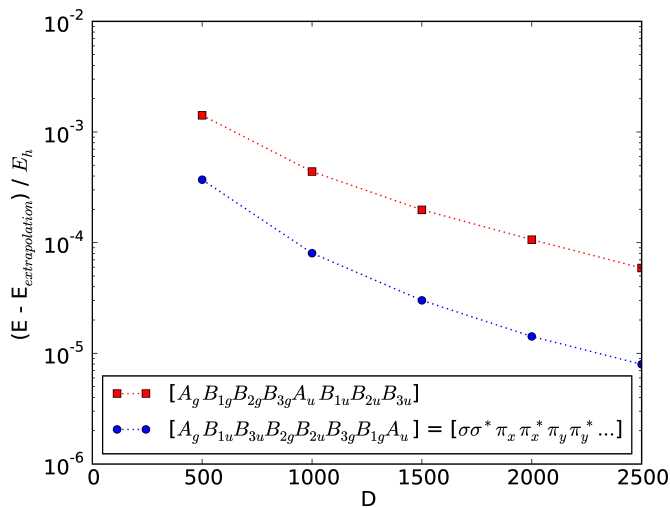}
\caption{\label{ConvergenceN2e-plot} The computational details for N$_2$ were discussed in the caption of Fig. \ref{ExtrapolN2-plot}. The energy converges significantly faster with the number of \textit{reduced} virtual basis states $D$ when the irrep blocks of bonding and anti-bonding molecular orbitals are placed next to each other.}
\end{figure}

Rissler, Noack and White proposed to use the two-orbital mutual information $I_{k,l}$ to order the orbitals \cite{Rissler2006519}. In addition to the single-orbital entropies $S_1(k)$ and $S_1(l)$, the two-orbital entropy $S_2(k,l)$ is also needed to calculate $I_{k,l}$. It can be obtained by choosing for subsystem $A$ the two orbitals $k$ and $l$. $S_2(k,l)$ can again be efficiently calculated in the DMRG algorithm, as its RDM can be built from expectation values of operators acting on at most two sites \cite{Rissler2006519}. The so-called subadditivity property of the entanglement entropy dictates that:
\begin{equation}
S_2(k,l) \leq S_1(k) + S_1(l).
\end{equation}
Any entanglement between orbitals $k$ and $l$ reduces $S_2(k,l)$ with respect to $S_1(k) + S_1(l)$. The two-orbital mutual information is defined by:
\begin{equation}
I_{k,l} = \frac{1}{2} \left( S_1(k) + S_1(l) - S_2(k,l) \right)(1 - \delta_{k,l}) \geq 0,
\end{equation}
and is thus a symmetric measure of the correlation between orbitals $k$ and $l$. Its bandwidth can be minimized, for example based on cost functions such as
\begin{equation}
I = \sum\limits_{k,l} I_{k,l} |k-l|^{\eta}.
\end{equation}
Rissler, Noack and White found no clear correspondence between $I_{k,l}$ and the integral measures of section \ref{sec-integral-measures}. They observed that $I_{k,l}$ is large between orbitals which belong to the same molecular point group irrep, as well as between corresponding bonding and anti-bonding orbitals with large partial occupations (far from empty or doubly occupied) \cite{Rissler2006519}. Later studies of various groups supported this finding and corresponding ordering \cite{kurashige:234114, yanai:024105, PhysRevA.83.012508, JPCLentanglement, ma:224105, 2013arXiv1312.2415W}. For small molecules such as dimers, it is best to group orbitals of the same molecular point group irrep into blocks, and place irrep blocks of bonding and anti-bonding type next to each other. If in addition natural orbitals (NO) are used, the orbitals within an irrep block should be reordered so that the ones with NO occupation number (NOON) closest to one, are nearest to the block of their bonding or anti-bonding colleagues \cite{ma:224105}. Fig. \ref{ConvergenceN2e-plot} illustrates the speed-up in energy convergence by reordering the point group irreps.

\section{Variations on QC-DMRG} \label{DMRG-algos}
\subsection{Quadratic scaling DMRG} \label{QS-DMRG-section}
For elongated molecules, when the active space is studied in a localized basis,
\begin{equation}
v_{ij;kl} = \int d\vec{r}_1 d\vec{r}_2 \frac{\phi^*_i(\vec{r}_1) \phi_k(\vec{r}_1) \phi^*_j(\vec{r}_2) \phi_l(\vec{r}_2)}{| \vec{r}_1 - \vec{r}_2 |} \label{physics-notation-two-body-matrix-elements}
\end{equation}
vanishes faster than exponential with the separation of orbitals $i$ and $k$, and the separation of orbitals $j$ and $l$. By defining a threshold, below which these two-body matrix elements can be neglected, one can reduce the cost of the QC-DMRG algorithm in Tab. \ref{compuReqDMRG} to $\mathcal{O}(L^2D^3)$ computational time, $\mathcal{O}(LD^2)$ memory, and $\mathcal{O}(L^2D^2)$ disk \cite{WhiteQCDMRG, hachmann:144101, dorando:084109}. Quadratic scaling DMRG (QS-DMRG) is not variational anymore because the Hamiltonian is altered, but the error can be controlled with the threshold. At present, QC-DMRG can achieve FCI energy accuracy for about 40 electrons in 40 highly correlated orbitals (in compact molecules) \cite{sharma:124121, 2013arXiv1312.2415W}. With QS-DMRG, one can achieve FCI energy accuracy for 100 electrons in 100 orbitals \cite{hachmann:144101}, and maybe more. It should however be repeated, that this method relies on the topology of the molecule, and exploits the fact that DMRG works very well for one-dimensional systems.

\subsection{Building-in dynamic correlation}
QC-DMRG can at present achieve FCI energy accuracy for about 40 electrons in 40 orbitals. The static correlation in active spaces up to this size can hence be resolved, while dynamic correlation has to be treated a posteriori. Luckily, QC-DMRG allows for an efficient extraction of the two-body RDM (2-RDM) \cite{zgid:144115, ghosh:144117}. The 2-RDM is not only required to calculate analytic nuclear gradients \cite{Chan2002, Spiropyran}, but also to compute the gradient and the Hessian in CASSCF \cite{Roos3}. It is therefore natural to introduce a CASSCF variant with DMRG as active space solver, DMRG-CASSCF or DMRG-SCF \cite{zgid:144116, ghosh:144117, ChanQUA:QUA22099}. Static correlation can be treated with DMRG-SCF. To add dynamic correlation as well, three methods have been introduced.

With more effort, the 3-RDM and some specific contracted 4-RDMs can be extracted from DMRG as well. These are required to apply second-order perturbation theory to a CASSCF wavefunction, called CASPT2, in internally contracted form. The DMRG variant is called DMRG-CASPT2 \cite{kurashige:094104,ma:224105,Spiropyran}.

Based on a CASSCF wavefunction, a configuration interaction expansion can be introduced, called MRCI. Recently, an internally contracted MRCI variant was proposed, which only requires the 4-RDM \cite{saitow:044118}. By approximating the 4-RDM with a cumulant reconstruction from lower-rank RDMs, DMRG-MRCI was made possible \cite{saitow:044118}.

Yet another way is to perform a canonical transformation (CT) on top of an MR wavefunction, in internally contracted form. When an MPS is used as MR wavefunction, the method is called DMRG-CT \cite{yanai:024105, neuscamman:024106, C2CP23767A}.

\subsection{Excited states} \label{DMRG-ExcitedStatesSectionInChatperTwo}
In addition to ground states, DMRG can also find excited states. By projecting out lower-lying eigenstates \cite{2013arXiv1312.2415W}, or by targeting a specific energy with the harmonic Davidson algorithm \cite{dorando:084109}, DMRG solves for a particular excited state. In these state-specific algorithms, the whole renormalized basis is used to represent one single eigenstate. In state-averaged DMRG, several eigenstates are targeted at once to prevent root-flipping. Their RDMs are weighted and summed to perform the DMRG renormalization step \cite{HallbergBook}. The renormalized basis then represents several eigenstates simultaneously.

DMRG linear response theory (DMRG-LRT) \cite{dorando:184111} allows to calculate response properties, as well as excited states. Once the ground state has been found, the MPS tangent vectors to this optimized point can be used as an (incomplete) variational basis to approximate excited states \cite{dorando:184111, PhysRevB.85.035130, PhysRevB.85.100408, PhysRevB.88.075122, PhysRevB.88.075133, NaokiLRTpaper}. As the tangent vectors to an optimized Slater determinant yield the configuration interaction with singles (CIS), also called the Tamm-Dancoff approximation (TDA), for HF theory \cite{helgaker2}, the same names are used for DMRG: DMRG-CIS or DMRG-TDA. The variational optimization in an (incomplete) basis of MPS tangent vectors can be extended to higher-order tangent spaces as well. DMRG-CISD, or DMRG configuration interaction with singles and doubles, is a variational approximation to target both ground and excited states in the space spanned by the MPS reference and its single and double tangent spaces \cite{PhysRevB.88.075122}.

By linearizing the time-dependent variational principle for MPS \cite{PhysRevLett.107.070601}, the DMRG random phase approximation (DMRG-RPA) is found \cite{2011arXiv1103.2155K, PhysRevB.88.075122, PhysRevB.88.075133, NaokiLRTpaper}, again in complete analogy with RPA for HF theory.

\subsection{Other ansatzes}
Two other related ansatzes have been employed in quantum chemistry: the TTNS \cite{PhysRevB.82.205105, nakatani:134113, LegezaTTNS} and the complete-graph TNS (CGTNS) \cite{1367-2630-12-10-103008, Marti:C0CP01883J}:
\begin{equation}
\ket{\Psi} = \sum\limits_{\{ n_k \}} \left( \prod\limits_{ i < j} C[i,j]^{n_i n_j} \right) \ket{n_1 ... n_L}. \label{CGTNS}
\end{equation}
The latter is an example of a correlator product state (CPS) \cite{1367-2630-11-8-083026}, in which multiple tensors can have the same physical index. The TTNS requires a smaller virtual dimension than DMRG to achieve the same accuracy. The accuracy of the CGTNS is limited by the number of correlated orbitals in each cluster (two in Eq. \eqref{CGTNS}). For a given desired accuracy, the optimization algorithms for TTNS and CGTNS are currently less efficient than QC-DMRG. As a result, QC-DMRG is still the preferred choice for \textit{ab initio} quantum chemistry.

There is also a QC-DMRG algorithm for the relativistic many-body four-component Dirac equation \cite{Knecht-4c-DMRG}.

\section{Symmetry} \label{SymmetrySection}

\subsection{Introduction}

The symmetry group of a Hamiltonian can be used to reduce the dimensionality of the exact diagonalization problem \cite{Weyl, Wigner1939}. The Hamiltonian does not connect states which belong to different irreps or to different rows of the same irrep. By choosing a basis of symmetry eigenvectors, the Hamiltonian becomes block diagonal, and each block can be diagonalized separately. The blocks which belong to different rows of the same irrep are closely related, and yield the same energies. In section \ref{entanglement-section}, it was discussed how locality leads to low-entanglement wavefunctions. These allow to reduce the dimensionality of the exact diagonalization problem as well, at least for ground and low-lying eigenstates. Symmetry and locality can be combined, which is shown in this section for DMRG.

From the very beginning, the abelian particle-number and spin-projection symmetries were incorporated in QC-DMRG \cite{WhiteQCDMRG,mitrushenkov:6815,Chan2002}. Abelian point group symmetry followed quickly \cite{chan:6110, PhysRevB.68.195116}. These symmetries are easy to implement, because they commute with the DMRG RDM. For $\mathsf{SU(2)}$ spin symmetry this is not the case, which is why its implementation took longer.

Sierra and Nishino first introduced exact $\mathsf{SU(2)}$ spin symmetry into DMRG with the interaction-round-a-face DMRG method \cite{Sierra1997505}. McCulloch and Gul\'acsi later found an easier way, based on a quasi-RDM \cite{McCulloch1, McCUlloch2, 0295-5075-57-6-852}, see section \ref{theo-symm-sec}. For the underlying MPS, this boils down to assuming that the rank-three MPS tensors are irreducible tensor operators of the symmetry group \cite{1742-5468-2007-10-P10014}. This opened the path to implement multiplicity-free non-Abelian symmetries also in TNSs \cite{1367-2630-12-3-033029, PhysRevA.82.050301, PhysRevB.86.195114}. The spin-adapted DMRG method of McCulloch and Gul\'acsi was later introduced in nuclear structure calculations \cite{PhysRevC.73.014301, PhysRevLett.97.110603, PhysRevC.78.041303}, where it is known as angular momentum DMRG or JDMRG, as well as in QC-DMRG \cite{zgid:014107, sharma:124121, woutersJCP1, 2013arXiv1312.2415W}. Non-multiplicity-free symmetries can also be exploited in DMRG, but require special considerations \cite{Weichselbaum20122972}.

Before the introduction of exact $\mathsf{SU(2)}$ symmetry in QC-DMRG, several tricks were employed. Legeza used a spin-reflection operator to distinguish even- and odd-spin states based on their spin parity \cite{PhysRevB.56.14449,PhysRevB.67.125114,DMRG_LiF}. A level shift operator \cite{moritz:184105, marti:014104, ghosh:144117,1367-2630-12-10-103008}
\begin{eqnarray}
\hat{H} & = & \hat{H}_0 + \alpha \hat{S}^- \hat{S}^+ \\
\hat{H} & = & \hat{H}_0 + \alpha \hat{S}^2 
\end{eqnarray}
can also be used to raise higher spin states in energy. Zgid and Nooijen \cite{zgid:014107} used the quasi-RDM to impose exact $\mathsf{SU(2)}$ spin symmetry in QC-DMRG, but they retained all states of a multiplet explicitly in the renormalized basis. In the works of Sharma and Chan \cite{sharma:124121} and Wouters \cite{woutersJCP1, 2013arXiv1312.2415W}, the Wigner-Eckart theorem was exploited to work with \textit{reduced} renormalized basis states instead of entire multiplets.

\subsection{The quasi-RDM method for $\mathsf{SU(2)}$ spin symmetry} \label{theo-symm-sec}
McCulloch's quasi-RDM method \cite{McCulloch1, McCUlloch2, 0295-5075-57-6-852, 1742-5468-2007-10-P10014} is reviewed in this section. Consider the bases $\{ \ket{j_A j^z_A \alpha_A} \}$ and $\{ \ket{j_B j^z_B \alpha_B} \}$ for subsystems $A$ and $B$ respectively, which have good spin $j$ and spin projection $j^z$ quantum numbers. $\alpha$ keeps track of the number of basis states with symmetry $(j,j^z)$. The wavefunction for the compound system with spin $S$ and spin projection $S^z$ can be written as
\begin{equation}
\ket{\Psi} = \sum\limits_{j_A j^z_A \alpha_A j_B j^z_B \alpha_B} \Psi^{S S^z}_{(j_A j^z_A \alpha_A) ; (j_B j^z_B \alpha_B)} \ket{j_A j^z_A \alpha_A} \ket{j_B j^z_B \alpha_B}. \label{McCullochWfn}
\end{equation}
The coefficients $\Psi^{S S^z}_{(j_A j^z_A \alpha_A) ; (j_B j^z_B \alpha_B)}$ are not completely independent, but are related to each other by Clebsch-Gordan coefficients. The triangle condition for angular momentum and the sum rule for spin projections have to be fulfilled for example:
\begin{eqnarray}
|j_A - j_B| & \leq & S \leq j_A + j_B, \label{triangleRelation}\\
j_A^z + j_B^z & = & S^z.
\end{eqnarray}
Only if the compound wavefunction is a spin singlet, $j_A$ and $j_B$ are constrained to be equal in the summation. This implies that the RDM $\hat{\rho}^A$ for subsystem $A$ is in general not block-diagonal with respect to $j_A$, except if $\ket{\Psi}$ is a singlet:
\begin{eqnarray}
& \hat{\rho}^A = \sum\limits_{j_A j^z_A \alpha_A \widetilde{j}_A \widetilde{\alpha}_A}  \ket{j_A j^z_A \alpha_A} \bra{\widetilde{j}_A j^z_A \widetilde{\alpha}_A} \nonumber \\
&  \left( \sum\limits_{j_B j^z_B \alpha_B} \Psi^{S S^z}_{(j_A j^z_A \alpha_A) ; (j_B j^z_B \alpha_B)} \Psi^{S S^z *}_{(\widetilde{j}_A j^z_A \widetilde{\alpha}_A) ; (j_B j^z_B \alpha_B)} \right) .
\end{eqnarray}
The eigenvectors of $\hat{\rho}^A$ will then not be spin eigenvectors. One way to obtain a renormalized basis of spin eigenvectors, is by using the quasi-RDM. It can be obtained from $\hat{\rho}^A$ by setting the off-diagonal blocks, which connect different spin symmetry sectors, to zero:
\begin{eqnarray}
& \hat{\rho}^A_{\text{quasi}} = \sum\limits_{j_A j^z_A \alpha_A \widetilde{\alpha}_A} \ket{j_A j^z_A \alpha_A}  \bra{j_A j^z_A \widetilde{\alpha}_A} \nonumber \\
& \left( \sum\limits_{j_B j^z_B \alpha_B} \Psi^{S S^z}_{(j_A j^z_A \alpha_A) ; (j_B j^z_B \alpha_B)} \Psi^{S S^z *}_{(j_A j^z_A \widetilde{\alpha}_A) ; (j_B j^z_B \alpha_B)} \right) .
\end{eqnarray}
The eigenvectors of $\hat{\rho}^A_{\text{quasi}}$ are spin eigenvectors, and their probability of occurrence in subsystem $A$ is given by the corresponding eigenvalues of $\hat{\rho}^A_{\text{quasi}}$ \cite{McCulloch1}. Quasi-RDMs can be constructed analogously for other non-Abelian symmetries as well.

A performance gain in memory and computer time can be obtained by working with reduced basis states. If for all multiplets $(j, \alpha)$, all spin projections $j^z$ are present, a Clebsch-Gordan coefficient can be factorized from the coefficient tensor in Eq. \eqref{McCullochWfn} due to the Wigner-Eckart theorem:
\begin{eqnarray}
& \ket{\Psi} = \sum\limits_{j_A j^z_A \alpha_A j_B j^z_B \alpha_B} \braket{j_A j_A^z j_B j_B^z \mid S S^z} \Psi^{S}_{(j_A \alpha_A) ; (j_B \alpha_B)} \nonumber \\
& \ket{j_A j^z_A \alpha_A} \ket{j_B j^z_B \alpha_B}, \label{FullWfnButRedCoeffTensor}
\end{eqnarray}
or in reduced form:
\begin{equation}
\left| \Ket{\Psi} \right. = \sum\limits_{j_A \alpha_A j_B \alpha_B} \Psi^{S}_{(j_A \alpha_A) ; (j_B \alpha_B)} \left| \ket{j_A \alpha_A} \right. \left|\ket{j_B \alpha_B}\right. . \label{reducedWfn}
\end{equation}

The DMRG renormalization tranformation to augment the left renormalized basis with one site (containing one spin) can analogously be written as
\begin{eqnarray}
& \ket{j_{i} j_{i}^z \alpha_{i}} = \sum\limits_{ j_{i-1} j_{i-1}^z \alpha_{i-1} s_i s^z_i} A[i]^{(s_i s_i^z)}_{ (j_{i-1} j_{i-1}^z \alpha_{i-1}) ; (j_{i} j_{i}^z \alpha_{i})} \nonumber \\
& \ket{j_{i-1} j_{i-1}^z \alpha_{i-1}} \ket{s_i s^z_i}, \label{gagjhasgipakxfhg395}
\end{eqnarray}
or in reduced form as
\begin{equation}
\left| \ket{j_{i} \alpha_{i}} \right. = \sum\limits_{ j_{i-1} \alpha_{i-1} s_i} T[i]^{(s_i)}_{ (j_{i-1} \alpha_{i-1}) ; (j_{i} \alpha_{i})} \left|\ket{j_{i-1} \alpha_{i-1}}\right. \left|\ket{s_i}\right. ,
\end{equation}
with
\begin{eqnarray}
& A[i]^{(s_i s_i^z)}_{ (j_{i-1} j_{i-1}^z \alpha_{i-1}) ; (j_{i} j_{i}^z \alpha_{i})} = \nonumber \\
& \braket{j_{i-1} j_{i-1}^z s_i s_i^z \mid j_i j_i^z} T[i]^{(s_i)}_{ (j_{i-1} \alpha_{i-1}) ; (j_{i} \alpha_{i})}. \label{CGfromMPS}
\end{eqnarray}
$A[i]^{(s_i)}$ can therefore be regarded as an irreducible tensor operator with spin $s_i$.

An extra performance gain can be achieved if the operators in the Hamiltonian are irreducible tensor operators of the imposed symmetry group. For spin systems, the following operators are an example:
\begin{equation}
\left( \hat{S}_{-1}^1 , \hat{S}_{0}^1, \hat{S}_{1}^1 \right) = \left( \frac{\hat{S}_x - i \hat{S}_y}{\sqrt{2}} , \hat{S}_z, - \frac{\hat{S}_x + i \hat{S}_y}{\sqrt{2}} \right).
\end{equation}
Due to the Wigner-Eckart theorem
\begin{equation}
\braket{s_1 s_1^z \mid \hat{S}^1_m \mid s_2 s_2^z} = \braket{s_1 \mid\mid \hat{S}^1 \mid\mid s_2} \braket{s_2 s_2^z 1 m \mid s_1 s_1^z}, \label{WEforOP}
\end{equation}
renormalized operators can be obtained in reduced form by recoupling the irreducible tensor operators and the reduced renormalized basis states. Formally this boils down to contracting the common multiplets of the Clebsch-Gordan coefficients in Eqs. \eqref{CGfromMPS} and \eqref{WEforOP}. The tensor product of irreducible tensor operators can also be obtained by working solely with reduced quantities \cite{1742-5468-2007-10-P10014}. Examples can be found in \cite{woutersJCP1, my_phd}.

For the coupling to spin $S$ in Eq. \eqref{reducedWfn}, all spin symmetry sectors $j_A$ and $j_B$ which comply with Eq. \eqref{triangleRelation} have to be taken into account. This strategy to form a spin-$S$ wavefunction is hence less efficient for larger values of $S$. One way to circumvent the large summation, is by adding a noninteracting site at the right end of the one-dimensional lattice, with spin $S$ \cite{0295-5075-57-6-852}. At the position of the current micro-iteration, one can then simply recouple to a singlet state. Sharma called this the singlet-embedding strategy \cite{sharma:124121}. In section \ref{sec-CheMPS2}, the singlet-embedding strategy will arise naturally based on Eqs. \eqref{gagjhasgipakxfhg395}-\eqref{CGfromMPS}.

Eq. \eqref{FullWfnButRedCoeffTensor} allows to explicitly target a specific symmetry sector of the Hamiltonian. The wavefunction is then always an exact eigenstate of $\hat{S}^2$, irrespective of the virtual dimension $D$. A singlet-triplet gap can then for example be obtained by two ground state calculations, instead of several excited state calculations. For the latter, spin mixing can occur, because working in the $S^z=0$ symmetry sector does not imply anything about $S$. Explicit measurement of $\hat{S}^2$, and its evolution with $D$, should then be used to discern the spin $S$.

Another advantage is the memory reduction. $A[i]$ contains $(2s_i+1)D^2$ variables. Due to the Clebsch-Gordan coefficients in Eq. \eqref{CGfromMPS}, it becomes block-sparse. Whenever a Clebsch-Gordan coefficient is zero, the corresponding MPS tensor block does not need to be allocated. In addition, the symmetry block $(j_{i-1} , j_{i})$ in $A[i]$ is represented in reduced form in $T[i]$. $D(j_i)$ \textit{reduced} renormalized basis states correspond in fact to $(2j_i+1)D(j_i)$ individual renormalized basis states. Next to block-sparsity, Eq. \eqref{CGfromMPS} hence also encompasses information compression. The block-sparsity and the compression result in faster contractions over common indices. Next to a memory advantage, there is hence also an advantage in computational time.

\subsection{Symmetries in ab initio quantum chemistry} \label{sec-CheMPS2}
In this section, $\mathsf{SU(2)}$ spin symmetry, $\mathsf{U(1)}$ particle-number symmetry, and the abelian point group symmetries $\mathsf{P}$ with real-valued character tables,
\begin{equation}
\mathsf{P} \in \{ \mathsf{C_1}, \mathsf{C_i}, \mathsf{C_2}, \mathsf{C_s}, \mathsf{D_2}, \mathsf{C_{2v}}, \mathsf{C_{2h}}, \mathsf{D_{2h}} \}, \label{PointGroupsInCheMPS2}
\end{equation}
will be discussed. Sharma has recently imposed non-abelian point group symmetry as well \cite{1.4867383}, but this is beyond the scope of this review. Because these abelian groups $\mathsf{P}$ all have real-valued character tables, the direct product of any irrep $I_j$ with itself gives the trivial irrep $I_0$:
\begin{equation}
\forall I_j: ~ I_j \otimes I_j = I_0.
\end{equation}
The physical basis states of orbital $k$ correspond to the following symmetry eigenstates:
\begin{eqnarray}
\ket{-} & \rightarrow & \ket{s=0; s^z=0; N=0; I=I_0} \\
\ket{\uparrow} & \rightarrow & \ket{s=\frac{1}{2}; s^z=\frac{1}{2}; N=1; I=I_k} \\
\ket{\downarrow} & \rightarrow & \ket{s=\frac{1}{2}; s^z=-\frac{1}{2}; N=1; I=I_k} \\
\ket{\uparrow\downarrow} & \rightarrow & \ket{s=0; s^z=0; N=2; I=I_0}.
\end{eqnarray}
The virtual basis states are also labeled by the quantum numbers of $\mathsf{SU(2)} \otimes \mathsf{U(1)} \otimes \mathsf{P}$:
\begin{equation}
\ket{\alpha} \rightarrow \ket{j j^z N I \alpha}.
\end{equation}
The equivalent of Eq. \eqref{CGfromMPS} is then
\begin{eqnarray}
A[i]_{(j_L j_L^z N_L I_L \alpha_L) ; (j_R j_R^z N_R I_R \alpha_R)}^{(s s^z N I)} = \braket{j_L j_L^z s s^z \mid j_R j_R^z} \nonumber \\
 \delta_{N_L + N, N_R} \delta_{I_L \otimes I, I_R} T[i]^{(s N I)}_{(j_L N_L I_L \alpha_L);(j_R N_R I_R \alpha_R)}. \label{CheMPS2_WE_MPS}
\end{eqnarray}
The $\mathsf{SU(2)}$, $\mathsf{U(1)}$, and $\mathsf{P}$ symmetries are locally imposed by their Clebsch-Gordan coefficients. These express nothing else than resp. local allowed spin recoupling, local particle number conservation, and local point group symmetry conservation. The index $\alpha$ keeps track of the number of reduced renormalized basis states with symmetry $(j, N, I)$. This equation again encompasses block-sparsity and information compression.

\begin{figure}[t!]
\centering
\includegraphics[width=0.45\textwidth]{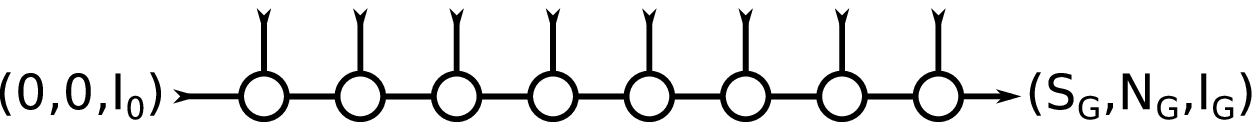}
\caption{\label{ImposingSymmetry-plot} Imposing $\mathsf{SU(2)}$, $\mathsf{U(1)}$, and $\mathsf{P}$ symmetry.}
\end{figure}

The desired global symmetry $(S_G,N_G,I_G)$ can be imposed with the singlet-embedding strategy, see Fig. \ref{ImposingSymmetry-plot}. Assume that the MPS is part of a larger DMRG chain, to which it is connected on its left and right ends. On the left end, there is only one irrep $(j_L,N_L,I_L) = (0,0,I_0)$ in the virtual bond, which has virtual dimension 1. On the right end, there is also only one irrep $(j_R,N_R,I_R) = (S_G,N_G,I_G)$ in the virtual bond, which also has \textit{reduced} virtual dimension 1. Eq. \eqref{CheMPS2_WE_MPS} and Fig. \ref{ImposingSymmetry-plot} imply that the addition of an extra orbital to the left renormalized basis is repeated from symmetry sector $(0,0,I_0)$ at boundary 0 to symmetry sector $(S_G,N_G,I_G)$ at boundary $L$.

\begin{figure}
\centering
\includegraphics[width=0.48\textwidth]{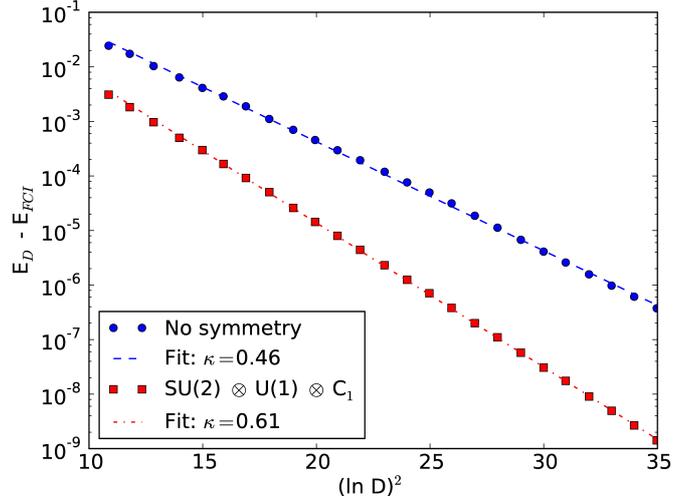} 
\caption{\label{ConvergenceHubbardSymmetry-plot} Convergence of the one-dimensional Hubbard model with open boundary conditions, $L=36$ sites, $N=22$ electrons, $U=6$, in the spin singlet state. The convergence scheme \eqref{Eextrapol2} is tested for a DMRG code without any imposed symmetries and for \textsc{CheMPS2} \cite{2013arXiv1312.2415W} with imposed $\mathsf{SU(2)} \otimes \mathsf{U(1)} \otimes \mathsf{C_1}$ symmetry. $\kappa$ is the parameter $C_4$ of Eq. \eqref{Eextrapol2}, and $D$ denotes the total number of renormalized basis states at each virtual bond. For \textsc{CheMPS2}, these are the \textit{reduced} ones.}
\end{figure}

Towards the middle of this embedded MPS chain, the reduced virtual dimension has to grow exponentially for the MPS to represent a general symmetry-adapted FCI state. To make the MPS ansatz in Eq. \eqref{CheMPS2_WE_MPS} of practical use, the total reduced virtual dimension per bond has to be truncated. The extrapolation scheme \eqref{Eextrapol2} is shown for the one-dimensional Hubbard model \cite{Hubbard26111963} with open boundary conditions
\begin{equation}
\hat{H} = - \sum\limits_{i=1}^{L-1} \sum\limits_{\sigma} \left( \hat{a}_{i\sigma}^{\dagger} \hat{a}_{i+1 \sigma} + \hat{a}_{i+1 \sigma}^{\dagger} \hat{a}_{i \sigma} \right) + U \sum\limits_{i=1}^{L} \hat{a}_{i\uparrow}^{\dagger} \hat{a}_{i\uparrow} \hat{a}_{i\downarrow}^{\dagger} \hat{a}_{i\downarrow}
\end{equation}
in Fig. \ref{ConvergenceHubbardSymmetry-plot}. The $\mathsf{SU(2)} \otimes \mathsf{U(1)} \otimes \mathsf{C_1}$ symmetry introduces block-sparsity and information compression. The latter can be seen in the faster energy convergence with the number of \textit{reduced} virtual basis states.

Due to the abelian point group symmetry $\mathsf{P}$, the matrix elements $h_{ij;kl}$ of the Hamiltonian \eqref{QC-ham-2} are only nonzero if $I_i \otimes I_j = I_k \otimes I_l$. If $\mathsf{P}$ is nontrivial, this considerably reduces the number of terms in the construction of the complementary renormalized operators, and in the multiplication of the effective Hamiltonian with a trial vector.

The operators
\begin{eqnarray}
\hat{b}^{\dagger}_{c \gamma} & = & \hat{a}^{\dagger}_{c \gamma} \label{creaannih1--}\\
\hat{b}_{c \gamma} & = & (-1)^{\frac{1}{2}-\gamma}\hat{a}_{c -\gamma} \label{creaannih2--}
\end{eqnarray}
for orbital $c$ correspond to resp. the $(s=\frac{1}{2}, s^z=\gamma, N=1, I_c)$ row of irrep $(s=\frac{1}{2}, N=1, I_c)$ and the $(s=\frac{1}{2}, s^z=\gamma, N=-1, I_c)$ row of irrep $(s=\frac{1}{2}, N=-1, I_c)$ \cite{BookDimitri}. $\hat{b}^{\dagger}$ and $\hat{b}$ are hence both doublet irreducible tensor operators. As described in section \ref{theo-symm-sec}, this fact permits exploitation of the Wigner-Eckart theorem for operators and (complementary) renormalized operators. Contracting terms of the type \eqref{CheMPS2_WE_MPS} and \eqref{creaannih1--}-\eqref{creaannih2--} can be done by implicitly summing over the common multiplets and recoupling the local, virtual and operator spins. As is shown in Refs. \cite{woutersJCP1, my_phd}, (complementary) renormalized operators then formally consist of terms containing Clebsch-Gordan coefficients and reduced tensors. In an actual implementation such as \textsc{Block} \cite{sharma:124121} or \textsc{CheMPS2} \cite{woutersJCP1, 2013arXiv1312.2415W}, only the reduced tensors need to be calculated, and Wigner 3-j symbols or Clebsch-Gordan coefficients are never used.

\section{QC-DMRG codes and studied systems} \label{DMRG-systems}
Table \ref{QC_DMRG-codes} gives an overview of the currently existing QC-DMRG codes. Two of them are freely available, \textsc{Block} and \textsc{CheMPS2}. Four codes have $\mathsf{SU(2)}$ spin symmetry: Zgid's code, \textsc{Rego}, \textsc{Block}, and \textsc{CheMPS2}. The former two explicitly retain entire multiplets at each virtual bond, while the latter two exploit the Wigner-Eckart theorem to work with a reduced renormalized basis and reduced renormalized operators, see section \ref{SymmetrySection}.

Two parallellization strategies are currently used: processes can become responsible of certain site indices of the (complementary) renormalized operators \cite{chan:3172}, or of certain symmetry blocks in the virtual bonds \cite{kurashige:234114}. For condensed-matter Hamiltonians, a real-space parallellization strategy has appeared recently \cite{PhysRevB.87.155137}, which might also be useful for QC-DMRG.

\begin{table}
\centering
\caption{\label{QC_DMRG-codes} Overview of QC-DMRG codes.}
\begin{tabular}{|l|l|l|}
\hline
Name                      & Authors             \\
\hline
                          & White               \cite{WhiteQCDMRG,Rissler2006519} \\
                          & Mitrushenkov        \cite{mitrushenkov:6815,QUA:QUA23173} \\
\textsc{Block}$^{(a)}$    & Chan \& Sharma      \cite{Chan2002,sharma:124121}\\
\textsc{Qc-Dmrg-Budapest} & Legeza              \cite{PhysRevB.67.125114,JCTCbondForm}\\
\textsc{Qc-Dmrg-Eth}      & Reiher              \cite{RichardsonControlReiher, JCTCspindens}\\
                          & Zgid                \cite{zgid:014107,zgid:144116}\\
                          & Xiang               \cite{PhysRevB.81.235129}\\
\textsc{Rego}             & Kurashige \& Yanai  \cite{kurashige:234114, naturechem}\\
\textsc{CheMPS2}$^{(b)}$  & Wouters             \cite{woutersJCP1,2013arXiv1312.2415W}\\
\textsc{Qc-Maquis}        & Keller \& Reiher    \cite{Accuracy-Keller}\\
\hline
\end{tabular}

$^{(a)}$\textsc{Block} is freely available from \cite{BlockCodeChan}.\\
$^{(b)}$\textsc{CheMPS2} is freely available from \cite{CheMPS2github}.
\end{table}

Many properties of many systems have been studied. QC-DMRG is of course able to calculate the ground state energy, but also excited state energies \cite{PhysRevB.67.125114, DMRG_LiF, moritz:184105, dorando:084109, ghosh:144117, Spiropyran, NaokiLRTpaper,2013arXiv1312.2415W, 1.4867383}, avoided crossings \cite{DMRG_LiF, moritz:184105, 2013arXiv1312.2415W, LegezaTTNS}, spin splittings \cite{hachmann:134309, marti:014104, zgid:014107, neuscamman:024106, mizukami:091101, 1367-2630-12-10-103008, sharma:124121, woutersJCP1, C3CP53975J, 2013arXiv1312.2415W, Harris2014, 10.1063.1.4885815}, polyradical character by means of the NOON spectrum \cite{hachmann:134309,ChanQUA:QUA22099,JCTCgrapheneNano}, static and dynamic polarizabilities \cite{dorando:184111, woutersJCP1}, static second hyperpolarizabilities \cite{woutersJCP1}, particle-particle, spin-spin, and singlet diradical correlation functions \cite{hachmann:134309,sharma:124121,JCTCgrapheneNano, saitow:044118}, as well as expectation values based on the 1- or 2-RDM such as spin densities \cite{JCTCspindens, newKura} and dipole moments \cite{DMRG_LiF}.

The systems which have been studied range from atoms and first-row dimers to large transition metal clusters and $\pi$-conjugated hydrocarbons. Several of them have repeatedly received attention in the QC-DMRG community:
\begin{itemize}
\item H$_2$O \cite{WhiteQCDMRG, Chan2002, PhysRevB.67.125114, chan:8551, PhysRevB.68.195116, chan:3172, PhysRevB.70.205118, RichardsonControlReiher, PhysRevB.81.235129, C2CP23767A, nakatani:134113, NaokiLRTpaper} was already the subject of several FCI studies, due to its natural abundance and small number of electrons.
\item Hydrogen chains \cite{Chan2002, hachmann:144101, zgid:144115, zgid:144116, QUA:QUA23173, woutersJCP1, nakatani:134113, ma:224105}: these one-dimensional systems exhibit large static correlation at stretched geometries. They are optimal testcases for QC-DMRG.
\item All-trans polyenes \cite{Chan2002, chan:204101, hachmann:144101, ghosh:144117, yanai:024105, saitow:044118, NaokiLRTpaper}: they are also one-dimensional, with a large MR character.
\item N$_2$ \cite{mitrushenkov:6815, Chan2002, mitrushenkov:4148, PhysRevB.68.195116, chan:6110, Rissler2006519, moritz:244109, C2CP23767A, nakatani:134113, JCTCbondForm, ma:224105, saitow:044118} was already the subject of several FCI studies, due to its MR character at stretched bond lengths and its small number of electrons.
\item Cr$_2$ \cite{mitrushenkov:6815, moritz:024107, moritz:034103, kurashige:234114, kurashige:094104, sharma:124121, nakatani:134113, ma:224105} is only found to be bonding at the CASPT2 level. A complete basis set extrapolation of DMRG-CASPT2 calculations in the cc-pwCV(T,Q,5)Z basis, correlating 12 electrons in 28 orbitals, was needed to retrieve an acceptable dissociation energy \cite{kurashige:094104}.
\item $\left[\text{Cu}_2\text{O}_2\right]^{2+}$ \cite{marti:014104, kurashige:234114, yanai:024105, PhysRevA.83.012508} requires accurate descriptions of both static and dynamic correlation along its isomerization coordinate. DMRG-CT, correlating 28 electrons in 32 orbitals, showed that the bis($\mu$-oxo) isomer is more stable than the $\mu-\eta^2:\eta^2$ peroxo isomer \cite{yanai:024105}.
\end{itemize}
Other QC-DMRG studies treat
\begin{itemize}
\item the avoided crossings in LiF \cite{DMRG_LiF, LegezaTTNS}, CsH \cite{moritz:184105, JCTCbondForm}, and C$_2$ \cite{2013arXiv1312.2415W}
\item the static correlation due to $\pi$-conjugation in acenes \cite{dorando:084109,hachmann:134309,JCTCgrapheneNano}, poly(phenyl) carbenes \cite{ChanQUA:QUA22099, mizukami:091101}, perylene \cite{C2CP23767A}, graphene nanoribbons \cite{JCTCgrapheneNano}, free base porphyrin \cite{neuscamman:024106, saitow:044118}, and spiropyran \cite{Spiropyran}
\item transition metal clusters such as $\left[\text{Fe}_2\text{S}_2(\text{SCH}_3)_4\right]^{2-}$ \cite{sharma:124121, NaokiLRTpaper}, $\left[\text{Fe(NO)}\right]^{2+}$ \cite{JCTCspindens, JPCLentanglement}, Mn$_4$CaO$_5$ in photosystem II \cite{naturechem}, the dinuclear oxo-bridged complexes $\left[\text{Fe}_2\text{O}\text{Cl}_6\right]^{2-}$ and $\left[\text{Cr}_2\text{O}(\text{NH}_3)_{10}\right]^{4+}$ \cite{Harris2014}, diferrate $\left[\text{H}_4\text{Fe}_2\text{O}_7\right]^{2+}$ \cite{C3CP55225J}, and oxo-Mn(Salen) \cite{10.1063.1.4885815}
\item molecules with heavy elements, for which relativistic effects become important, such as CsH \cite{moritz:184105, JCTCbondForm}, the complexation of CUO with four Ne or Ar atoms \cite{C3CP53975J}, and the binding energy of TlH \cite{Knecht-4c-DMRG}
\end{itemize}
For transition metal clusters, QC-DMRG is currently the only viable choice due to the large active spaces which have to be handled.

\section{Conclusion}
The DMRG algorithm is well understood by means of the underlying MPS wavefunction. This allows to assess DMRG with concepts from quantum information theory. Accurate extrapolation schemes are known for the evolution of the variational energy with increasing virtual dimension D, or with decreasing discarded weight. The use of symmetry to reduce the computational cost is also well understood. Most progress can still be made in the orbital choice and ordering for nontrivial orbital topologies.

The 2-RDM can be extracted efficiently from QC-DMRG, and is required to calculate the gradient and the Hessian in CASSCF. QC-DMRG is therefore an ideal candidate to replace the FCI solver in CASSCF. DMRG-SCF, as the method is called, can resolve the static correlation in active spaces of up to 40 electrons in 40 orbitals. Several dynamical correlation theories for CASSCF have been used with DMRG-SCF as well: DMRG-CASPT2, DMRG-MRCI, and DMRG-CT. QC-DMRG has not only the ability to provide accurate reference data, but for a number of challenging systems it is currently also the only viable choice. These features have made DMRG increasingly important for \textit{ab initio} quantum chemistry during the past 15 years, and undoubtedly the method will be indispensable in future years as well.

\section*{Acknowledgements}
Sebastian Wouters is grateful for a Ph.D. fellowship from the Research Foundation Flanders. The computational resources (Stevin Supercomputer Infrastructure) and services used in this work were provided by the VSC (Flemish Supercomputer Center), funded by Ghent University, the Hercules Foundation and the Flemish Government - department EWI.

\bibliographystyle{unsrtnat}
\bibliography{mybiblio}
%
%
%

\end{document}